# Layer-Dependent Electronic and Optical Properties of 2D Black Phosphorus: Fundamentals and Engineering

*Guowei Zhang,\* Shenyang Huang, Fanjie Wang, and Hugen Yan\**

Prof. G. Zhang
Institute of Flexible Electronics
Northwestern Polytechnical University
Xi'an 710072, Shaanxi, China
E-mail: iamgwzhang@nwpu.edu.cn

Dr. S. Huang, Dr. F. Wang, Prof. H. Yan
State Key Laboratory of Surface Physics
Department of Physics
and Key Laboratory of Micro and Nano Photonic Structures (MOE)
Fudan University
Shanghai 200433, China
E-mail: hgyan@fudan.edu.cn

**Abstract:**

In 2D materials, the quantum confinement and van der Waals-type interlayer interactions largely govern the fundamental electronic and optical properties, and the dielectric screening plays a dominant role in the excitonic properties. This suggests strongly layer-dependent properties, and a central topic is to characterize and control the interlayer interactions in 2D materials and heterostructures. Black phosphorus is an emerging 2D semiconductor with unusually strong interlayer interactions and widely tunable direct bandgaps from the monolayer to the bulk, offering us an ideal platform to probe the layer-dependent properties and the crossover from 2D to 3D (i.e., the scaling effects). In this review, we present a comprehensive and thorough summary of the fundamental physical properties of black phosphorus, with a special focus on the layer-dependence characters, including the electronic band structures, optical absorption and photoluminescence, and excitonic properties, as well as the band structure engineering by means of electrical gating, strain, and electrochemical intercalation. Finally, we give an outlook for the future research.

# 1. Introduction

In layered materials, each layer can be viewed as a basic building block, and the layers are held together by weak van der Waals (vdWs) forces, this enables the extraction of one or a few layers out of the parent materials, individually existing as two-dimensional (2D) materials with atomic thickness. In 2004, Novoselov and Geim et al. successfully isolated a single layer of carbon atoms (namely graphene) from graphite by mechanical exfoliation, becoming the first 2D materials.[1] Soon after that, 2D materials have attracted increasing attention in multi-disciplinary fields, including physics, chemistry, materials science and electronics.

Nowadays, a variety of exfoliated 2D materials have been discovered with diverse electronic properties, including semimetals (graphene),[1-6] insulators ($h$-BN),[7] ferromagnets ($Cr_2Ge_2Te_6$, $CrI_3$),[8,9] charge density wave (CDW) materials ($2H$-$NbSe_2$, $1T$-$TiSe_2$),[10,11] high-temperature superconductors ($Bi_2Sr_2CaCu_2O_{8+\delta}$),[12] and a large class of semiconductors,[13-23] such as black phosphorus (BP),[24] transition metal dichalcogenides (TMDCs: $MoS_2$, $MoSe_2$, $WS_2$, $WSe_2$),[25-28] group-IV monochalcogenides (SnS, SnSe, GeS, GeSe), group-III monochalcogenides (GaS, GaSe, InSe), et al. In 2D materials, interlayer interactions (or vdWs interactions) play a key role in defining the electronic properties, implying a lot of fascinating layer-tunable properties. For example, monolayer $CrI_3$ exhibits an out-of-plane ferromagnetic order, while interlayer interactions induce an antiferromagnetic behavior in bilayer $CrI_3$.[9] When TMDCs are thinned down to monolayer, an indirect-to-direct bandgap transition occurs with significantly enhanced photoluminescence (PL).[25,26] Besides, interlayer interactions are also strongly modified by stacking orders even for the same layer number $N$. This is manifested by the distinctly different electronic structures in few-layer graphene with Bernal (AB) and rhombohedral (ABC) stackings.[29,30] Most strikingly, interlayer interactions can be also artificially tuned through a twist angle between the two constituting layers. A lot of fantastic quantum phenomena have been observed in this new-type of vdWs heterostructures, such as unconventional superconductivity[31] and Mott-like insulator states[32] in magic-angle-twisted bilayer graphene, and moiré excitons in small-angle-twisted TDMCs/TMDCs.[33-36]

BP is a rising 2D material with unusually strong interlayer interactions. It was first discovered in the bulk form by Bridgman in 1914, transformed from white phosphorus at high pressure (1.2 GPa) and moderate temperature (200 ℃).[37] Right 100 years after that, Li et al. reported transistors made of exfoliated few-layer BP down to several nanometers, with the carrier mobility up to ~1000 $cm^2$ $V^{-1}$ $s^{-1}$ and on/off ratio of $10^5$ at room temperature.[24] The excellent device performance soon drew a renewed interest in BP as a new member of the 2D material library. Similar to graphite, BP has a layered structure composed of P atoms, with an interlayer distance of ~5.25 Å.[38] In each BP layer, six P atoms are covalently bonded to each other to form the hexagonal ring (highlighted in **Figure 1**a). Different from the planar structure of graphene, BP exhibits a puckered structure. In a single layer, P atoms are not arranged in the same plane, some are in the upper sublayer, and the others are in the lower sublayer, together constituting the unique puckered structure. The distance between the upper and lower sublayers is ~2.1 Å (Figure 1b).[38] BP belongs to the orthorhombic system with reduced symmetry,[23] in the layer plane, there are two characteristic crystal orientations: armchair (AC, parallel to the pucker) and zigzag (ZZ, perpendicular to the pucker), respectively (Figure 1a). This unique in-plane structural anisotropy leads to a lot of intrinsic anisotropic electrical,[39-41] optical,[40,42-46] thermal[47-49] and mechanical[50,51] properties, distinctly distinguishing BP from isotropic 2D materials.

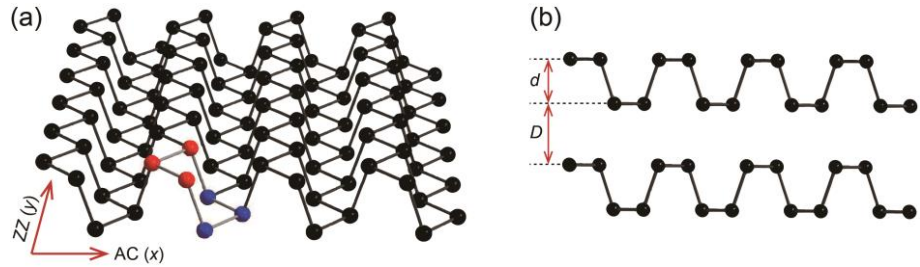

**Figure 1.** Structure of BP. a) Monolayer BP: one hexagonal ring composed of six P atoms is highlighted, with three in the upper sublayer (red) and the other three in the lower sublayer (blue). The two arrows indicate the AC and ZZ directions, respectively. b) Bilayer BP: side view along the AC direction. The interlayer ($d + D$) and inter-sublayer ($d$) distances are shown, respectively.

In addition, BP exhibits a direct bandgap for each layer number.[52,53] Due to strong interlayer interactions, the optical bandgap of BP can be tuned in a wide energy range, spanning from 1.7 eV (monolayer) to 0.3 eV (bulk),[42,43] bridging large-gap TMDCs and zero-gap graphene. Owing to the direct-gap nature, strong PL has been observed in BP with significant layer-dependence, covering a broad frequency range from visible to mid-IR.[43,54-56] In 2D BP, excitons are expected to dominate the optical properties, with the binding energy strongly depending on the number of layers.[57,58]

Theory predicts that other many-body quasiparticles, such as trions and biexcitons, hold significant layer-dependence as well.[59] For BP (except the monolayer), the optical response associated with the bandgap transition lies in the IR range, which can be further extended into longer IR wavelength by electric fields[60,61] and strain.[42,62,63] Combined with the relatively high carrier mobility, BP displays great potentials in high-speed and polarized IR devices, such as photodiodes,[64-67] optical modulators and lasers.[68]

In 2D materials, the number of layers can be easily tuned with atomic layer precision, along with the change in the quantum confinement, the dielectric environment and interlayer interactions, and hence to modify the material properties. For years, extensive efforts have been devoted to the characterization and control of interlayer interactions in 2D materials and heterostructures. BP has significant layer-dependent and direct bandgaps from the monolayer to the bulk, establishing it an ideal material system for study of layer-dependent properties from 2D to 3D (i.e., the scaling effects). This review article is aimed to systematically summarize the fundamental electronic and optical properties of BP, with a special focus on the layer-dependence characters, as well as the strategies for the band structure engineering. It is organized as follows: In Section 1, we start with a brief introduction of BP. In Section 2, we give a theoretical overview of the band structure of few-layer BP. Section 3 summarizes the recent experimental advances in optical absorption and PL of few-layer BP, particularly emphasizing the significant layer-dependence character. Section 4 presents the experimental progress in the band structure engineering of few-layer BP. In Section 5, we discuss the excitonic effects in few-layer BP mainly in the experimental side. In Section 6, we present a brief summary and an outlook for the future research.

## 2. Layer-Dependent Band Structure

For a semiconductor, the band structure largely determines the electronic and optical properties. For a long time after the discovery of BP in 1914, people know little about its band structure. Until 1953, the bandgap of bulk BP was determined to be 0.33 eV by electrical transport measurements. Besides, Hall measurements showed that BP is a *p*-type semiconductor due to impurities, especially at low temperature.[69] In 1963, Warschauer preformed IR absorption measurements for bulk BP, a clear absorption edge at ~4 μm (0.31 eV) was observed.[70] In 1981, theoretical calculations uncovered more information about the band structure of BP for the first time by Takao and Morita using a tight-binding (TB) method.[71] They found that monolayer BP is a direct bandgap semiconductor, with the conduction band minimum (CBM) and valence band maximum (VBM) both locating at Γ point of the 2D Brillouin zone. Besides, both of the wavefunctions of the lowest conduction band (CB) and the highest valence band (VB) are mostly composed of $3p_z$ component. For bulk BP, it exhibits a direct bandgap as well. However, unlike the monolayer case, interlayer couplings reduce (lift) the CB (VB) at Z point in the 3D Brillouin zone, making it lower (higher) than that at Γ point. As a result, the CBM and VBM both shift to Z point. The bandgap for bulk BP is calculated to be 0.33 eV, in excellent agreement with previous electrical[69] and optical[70] measurements.

Since 2014, extensive efforts have been devoted to the theoretical studies of the band structures in 2D BP,[52,53,72,73] stemming from the renewed boom of interest in BP as a new 2D member.[24,39,40,74] In this section, we will give an overview of the band structure of BP from the monolayer to the bulk, highlighting the features of anisotropy, band splitting and strong layer-dependence. Below, two representative work will be presented. One is the density functional theory (DFT) calculations by Qiao et al.,[52] in which bulk BP shows a direct bandgap (0.31-0.36 eV) at Z point, confirmed by recent angle-resolved photoemission spectroscopy (ARPES) measurements (**Figure 2**e).[43,75] However, direct ARPES measurements for 2D BP is rather lacking. As bulk BP thins down to monolayer, the direct bandgap still holds, but shifts to Γ point with a significantly increased value. Besides, the CB and VB both show strong anisotropy, which are very steep along the Γ-X direction (corresponding to the AC direction in real space), while nearly flat along the Γ-Y direction (the ZZ direction). This suggests light effective masses for electrons and holes in the AC direction, and heavy ones in the ZZ direction. In other words, the carriers transport faster in the AC direction than in the ZZ direction, which has been verified by experiments.[76,77]

In bilayer BP, interlayer coupling is introduced, the direct bandgap at Γ point undergoes a sharp reduce compared with monolayer. In *N*-layer (*N*L) BP, the bandgap is further decreased due to strong interlayer couplings. Most interestingly, the CB and VB spilt into *N* pairs of subbands, akin to transitional quantum wells (QWs), and these split conduction (and valence) subbands share the same origin in terms of the atomic orbitals. Moreover, the authors also predict moderately large mobilities for few-layer BP, on the order of $10^3$ cm$^2$ V$^{-1}$ s$^{-1}$. Recently, Long et al. reported an ultrahigh hole mobility of 5200 cm$^2$ V$^{-1}$ s$^{-1}$ in high-quality BP sandwiched between two *h*-BN flakes at room temperature, further increased to 45000 cm$^2$ V$^{-1}$ s$^{-1}$ at cryogenic temperatures.[78] The high mobility makes BP another 2D candidate beyond graphene for hosting fascinating quantum phenomena, such as quantum oscillations[79,80] and quantum Hall effects (QHE).[78,81-84]

The other is the tight-binding calculations by Rudenko et al.,[53,85] it is more clearly evidenced for the important role of interlayer couplings (or hoppings) in the band structures of BP, which is responsible for the band splitting and significant layer-dependence. As illustrated in Figure 2a, ten intralayer and four interlayer hopping parameters are involved in the calculations, among which three are the most important and dominant, i.e., two intralayer hoppings ($t_\parallel^1$ and $t_\parallel^2$, $t_\parallel^1 < 0$, $t_\parallel^2 > 0$), and one interlayer hopping ($t_\perp$), the former two determine the bandgap of monolayer BP, while the last plays a crucial role in the few-layer case. Figures 2b-d present the band structures for monolayer, bilayer and trilayer BP obtained by the TB calculations, respectively. It clearly shows that the bandgaps are all direct at Γ point. In monolayer, $t_\parallel^2$ has a larger absolute value over $t_\parallel^1$, playing a dominant role in determining the bandgap value. In bilayer, the CB and VB both split into two subbands, as shown in Figure 2c, labelled as CB1 and CB2, VB1 and VB2, respectively. With respect to the CB in monolayer, interlayer couplings bring down the CB1 and lift the CB2 in bilayer, similar trends are also applied to the VBs, thus the fundamental bandgap between CB1 and VB1 at Γ point is reduced.

The authors found that the interlayer hopping leads to this splitting, and the energy separation between CBs (and VBs) is proportional to $t_\perp$. From the monolayer to the bulk, the quasiparticle bandgap decreases from 2.0 eV to 0.3 eV, the tuning range is much larger than most of other 2D materials, this significant layer-tunability originates from the strong interlayer interactions. As seen from Figure 1b, the puckered structure leads to the fact that the effective interlayer distance (i.e., the distance *D* between the upper sublayer of the bottom layer and the lower sublayer of the top layer) is smaller than that of other 2D materials, and hence the interlayer coupling is much stronger. The ultralow-frequency Raman experiment on few-layer BP provides a direct evidence for the unusually strong interlayer couplings.[86]

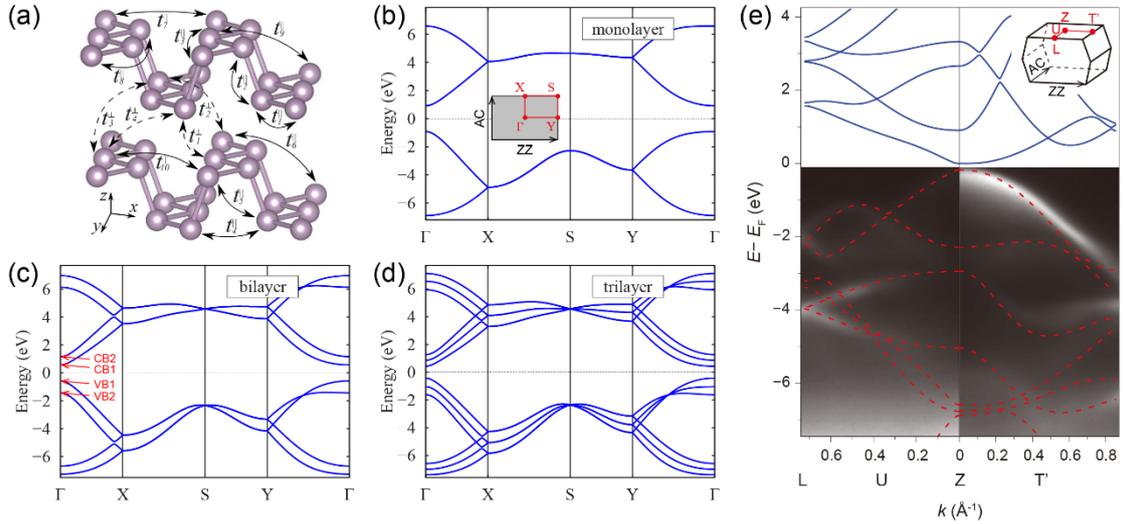

**Figure 2.** Band structure of BP. a) Schematic of the intralayer and interlayer hopping parameters in the tight-binding model. b-d) Calculated band structures of monolayer, bilayer and trilayer BP, respectively. The 2D Brillouin zone is illustrated in the inset of Figure 2b. Figures 2a-d are Reproduced with permission.[85] Copyright 2015, American Physical Society. e) Band structure of bulk BP obtained by ARPES measurements, with the calculated bands (solid and dashed lines) superimposed on its top. The 3D Brillouin zone is illustrated in the inset. Reproduced with permission.[24] Copyright 2014, Springer Nature.

Experimentally, the subband structures in few-layer BP have been confirmed by low-temperature scanning tunneling microscopy (LT-STM) studies.[87] **Figure 3**c shows a typical d$I$/d$V$ spectrum of few-layer BP, which reflects the density of state (DOS). As indicated by the orange arrows in CB and green arrows in VB, a set of well-resolved resonant features are clearly observable. The first and second pairs of subbands are clearly resolved, denoted as $C_1$ and $V_1$, and $C_2$ and $V_2$, respectively.

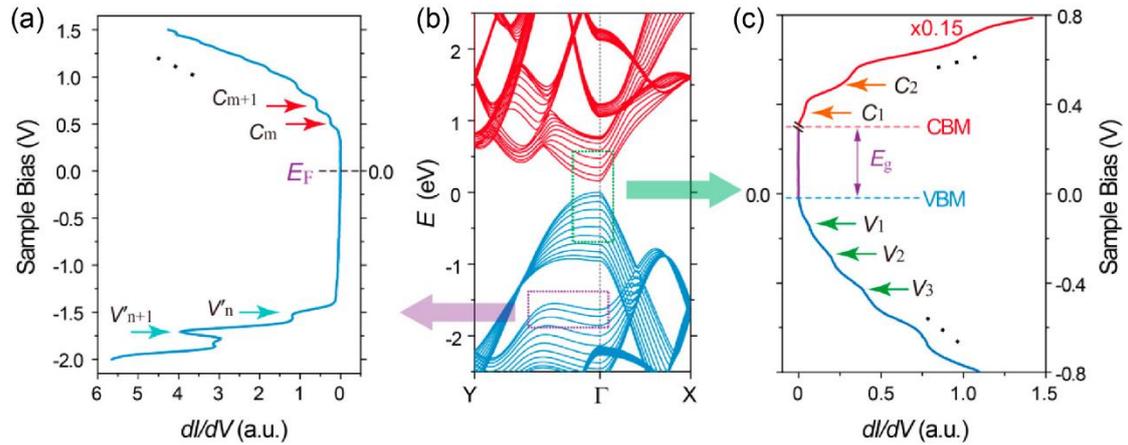

**Figure 3.** STM study of few-layer BP. a) A wide energy range d$I$/d$V$ spectrum of few-layer BP. b) Theoretically calculated band structure of 11L BP. c) A narrow energy range d$I$/d$V$ spectrum of the same BP in Figure 3a, corresponding to the low-lying bands (marked by a green box in Figure 3b). Reproduced with permission.[87] Copyright 2017, American Chemical Society.

## 3. Layer-Dependent Optical Properties

Tunable optical properties of materials are highly desirable in optoelectronics and photonics. In thick BP, nonlinear optical absorption[88] and third-harmonic generation (THG)[89-91] have been demonstrated to be thickness-dependent. Raman spectroscopy is widely used to determine the layer number of graphene[92] and MoS$_2$.[93] While for BP, the Raman behavior ($A_g^1$, $B_{2g}$, $A_g^2$) is rather complicated, depending on many factors (e.g., the sample thickness, the excitation wavelength, Raman modes, polarization angles).[94-100] The low-frequency Raman modes, which are sensitive to interlayer couplings in layered materials, show strong layer-dependence in few-layer BP.[86] In the time-domain, Miao et al. demonstrated that coherent phonon dynamics in few-layer BP is strongly layer-dependent as well.[101] Detailed Raman properties of BP have been reviewed elsewhere.[102] In this section, we focus on the linear optical absorption and PL of BP in the experimental side, emphasizing the significant layer-dependence character. First, we start with a brief introduction of the production of 2D BP, followed by a detailed discussion on anisotropic and layer-dependent optical absorption, as well as the strong PL observed in BP from the monolayer to the bulk.

### 3.1. Mechanical Exfoliation and Optical Contrast

For practical device applications, the large-scale synthesis of 2D BP is highly desirable. At current stage, it is quite challenging and there are very few attempts, with the strategy of phase transition from red phosphorus films to BP films on a substrate at proper pressure/temperature conditions. Xia group at Yale University reported wafer-scale synthesis of polycrystalline BP thin films (several tens of nanometers in thickness) on polyethylene terephthalate (PET) or sapphire substrates, with the maximum value of carrier mobility reaching ~160 cm$^2$ V$^{-1}$ s$^{-1}$ at room temperature.[103,104] Smith et al. reported the direct growth of BP films on Si substrates with areas >100 μm$^2$, and few-layer BP (~4L) with areas >3 μm$^2$.[105] Up to date, mechanical exfoliation is still the most commonly used technique for preparing high-quality 2D BP. Nevertheless, only small thick flakes can be obtained by the conventional method,[1] since BP is not so cleavable as graphite. To overcome this issue, an improved method was developed with the assistance of polydimethylsiloxane (PDMS).[106] In brief, as illustrated in **Figure 4**a, bulk BP is first cleaved several times by the Scotch tape. Next, the tape containing micro-crystals of BP is slightly pressed against a piece of PDMS, and then peeled off quickly. Some thin BP flakes with reasonable size and clean surface will be left on the PDMS, which are easily to be dry transferred to other substrates. Figure 4b shows an optical image of an exfoliated 2L BP on PDMS, with a large size of 100 μm × 290 μm.

The layer number of 2D BP can be estimated according to the optical contrast (OC), this method is widely applied to other 2D materials.[107,108] Figure 4c shows an exfoliated BP flake containing 1-4L on PDMS, with different regions clearly resolved by different colors. Figure 4d presents the OC (in the green CCD channel) along the two line cuts in Figure 4c in the reflection mode. It is found that the OC exhibits a step-like feature, one-to-one corresponding to each layer number. With per layer added, the OC increases by ~20%.[42] The layer-dependent OC provides a very simple way to estimate the layer number of 2D BP, and accurate determination is achieved by IR characterization, which will be discussed in detail later. It should be noted that i) this method is only suitable for thin layers (possibly <10L), since the OC for thick layers is not so distinguishable. ii) the step height for OC differs with substrates. For $SiO_2$/Si substrate, the OC (in the red CCD channel) only increases by ~7% with per

BP layer added.[43] iii) the OC can be largely enhanced by optical interferences, which can improve the discrimination for adjacent layer thickness.[56]

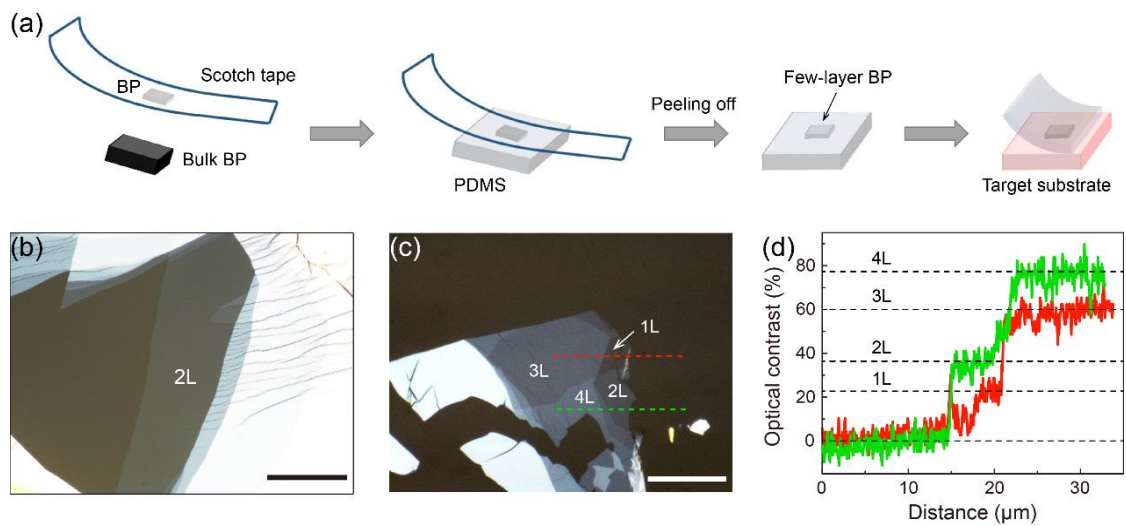

**Figure 4.** a) Schematic of PDMS-assisted mechanical exfoliation. b) Optical image of a large-area 2L BP on PDMS, the scale bar is 100 μm. c) Optical image of a BP flake on PDMS, containing 1-4L. The scale bar is 20 μm. d) Optical contrast in the green CCD channel along the two line cuts in Figure 4c. Figures 4c-d are Reproduced with permission.[42] Copyright 2017, Springer Nature.

### 3.2. Anisotropic Optical Absorption

Strong optical anisotropy has been observed in monolayer, few-layer and bulk BP.[40,42,43,57] Taking an 8L BP[57] as an example, **Figure 5**a shows a typical IR extinction ($-\Delta T/T_0$) spectrum for few-layer BP, with different light polarizations from 0° to 90°, in a step of 15°, where $T$ and $T_0$ denote light transmittance for samples on substrate and bare substrate, respectively (**Figure 6**e). The incident IR beam is normal to the layer plane. For 2D materials on thick and transparent substrates (the thickness of the substrate is far larger than the wavelength of incident light, thereby the interference effect is ignored), if the absorption is not large, the extinction ($-\Delta T/T_0$) is approximately in proportion to (the real part of) the optical conductivity (or absorption).[109,110]

For the 0° polarization, two sharp peaks appear in the spectrum, labelled as $E_{11}$ and $E_{22}$, respectively. The optical transitions are illustrated in Figure 5c, $E_{11}$ denotes the optical transition from VB1 to CB1, and $E_{22}$ is from VB2 to CB2, respectively. Few-layer BP can be viewed as an infinite QW, with the air on its top and the substrate beneath it acting as infinite deep barriers. In traditional QWs, for normal light incidence, only $E_{jj}$ transitions with the same subband index ($\Delta j = 0$) are allowed.[45,111] For the 8L case, eight branches of optical transitions ($E_{11}$ to $E_{88}$) should be observable, yet the energies of higher subband transitions are out of the measurement range. Note that in low-dimensional materials, excitonic effects dominate the optical response, due to significantly reduced dielectric screening and enhanced quantum confinement, as has been confirmed in 1D carbon nanotubes[112,113] and 2D TMDCs.[114-117] Thus, the observed sharp peaks are due to excitons, with the resonance energy below the quasiparticle bandgap. Even though, optical transitions are largely determined by the band structure, thus in this paper (except Section 5 focusing on excitons), the optical transition is interpreted in the single-particle picture.

As the light polarization changes from 0° to 90°, the $E_{11}$ and $E_{22}$ intensities decrease from the maximum to the minimum, showing strong in-plane anisotropy. This is originated from the band anisotropy.[45,52] In the ZZ direction, the mirror-symmetry strictly forbids the optical transition at Γ point (or Z point in 3D), while it is allowed in the AC direction (detailed discussions are available in Refs.[118,119], and more optical anisotropy, e.g., phonon absorption and free-carrier absorption, are reviewed elsewhere[120]). The polarization dependences of IR extinction at $E_{11}$ and $E_{22}$ in the range of 0° to 360° are summarized in Figure 5b, a two-fold pattern is clearly seen for both transitions. For the AC polarization (0°/180°), BP absorbs the most light, while for the ZZ polarization (90°/270°), the absorption is nearly zero.[57] For BP with other index $(N, j)$, the $E_{jj}^N$ transition keeps the same anisotropy.[42] This unique polarization dependence provides a simple and reliable way to determine the crystal orientation of BP.

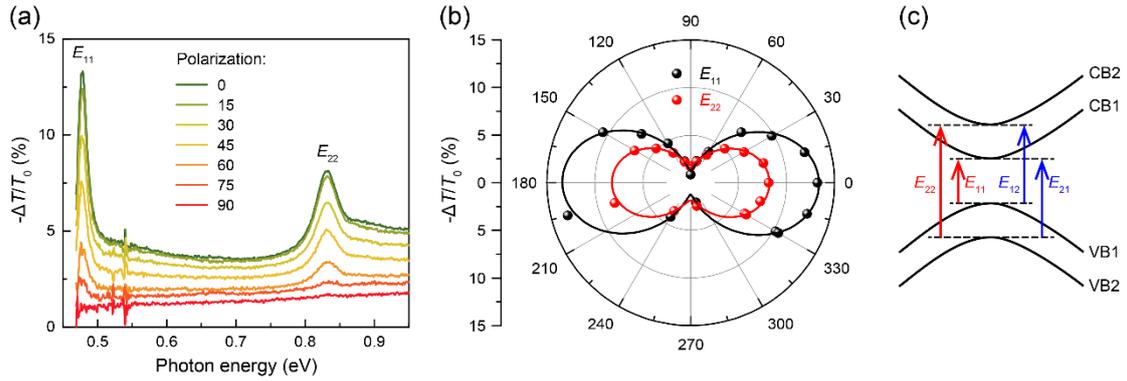

**Figure 5.** Anisotropic absorption in BP. a) IR extinction ($-\Delta T/T_0$) spectra of 8L BP on PDMS substrate, with different light polarizations from 0° to 90°. b) IR extinction at $E_{11}$ (black) and $E_{22}$ (red) as a function of the polarization angle $\theta$, the solid lines are $\cos^2\theta$ fits. Figures 5a-b are Reproduced with permission.[57] Copyright 2018, American Association for the Advancement of Science. c) Schematic illustration of optical transitions between quantized subbands, $E_{11}$ and $E_{22}$ denote (allowed) main transitions with the same subband index ($\Delta j = 0$), and $E_{12}$ and $E_{21}$ denote (forbidden) hybrid transitions with different subband index ($\Delta j \neq 0$), respectively.

### 3.3. Layer-Dependent Optical Absorption

The most compelling aspect for BP is the significant layer-tunability. Li et al. reported the first optical absorption study of 1-5L BP in the visible to near-IR range (0.75-2.5 eV), with the bandgap transition ($E_{11}$) only observable for 1-3L.[43] To overcome this limitation, Zhang et al. presented a comprehensive IR absorption study of BP from 2L to bulk, within a wide energy range (0.1-1.36 eV), using Fourier transform infrared spectroscopy (FTIR).[42] For the completeness, the data of 1L is also shown here. Figure 6a summarizes the reflectance contrast ($-\Delta R/R_0$) spectrum for 1L on $SiO_2$/Si substrate and IR extinction ($-\Delta T/T_0$) spectra for 2-11L on PDMS substrate. Unless stated otherwise in this paper, the polarization of incident light is along the AC direction. Note that the spectrum for 1L is modified by optical interferences from the $SiO_2$ layer (300 nm), thus it makes no sense to compare the absolute intensity between 1L and few-layer BP.

As can be seen, a sharp peak is clearly observed for 1L BP at ~1.69 eV, assigned to the bandgap transition ($E_{11}$), in excellent agreement with Li's results.[43] As for 2L, the $E_{11}$ peak shifts to 1.11 eV, with a significant decrease. This is originated from the strong interlayer couplings.[52,53,72] As the layer number further increases, the $E_{11}$ monotonically redshifts. Meanwhile, additional peaks ($E_{22}$, $E_{33}$, …), originated from higher subband transitions, appear in the spectra within the FTIR measurement range, which redshift with increasing layer number as well. For thicker ones (e.g., 13L and 15L, shown in **Figure 7**a), at least four peaks are well-resolved, the energy spacing between adjacent $E_{jj}$ (e.g., $E_{22} - E_{11}$, $E_{33} - E_{22}$, et al.) decreases with layer number. In the bulk limit, the split CBs (and VBs) are so close to each other that they are not distinguishable at all. As a consequence, they eventually evolve to a quasi-continuous band, this is manifested by the vanishing subband transitions in bulk BP (Figure 6c).

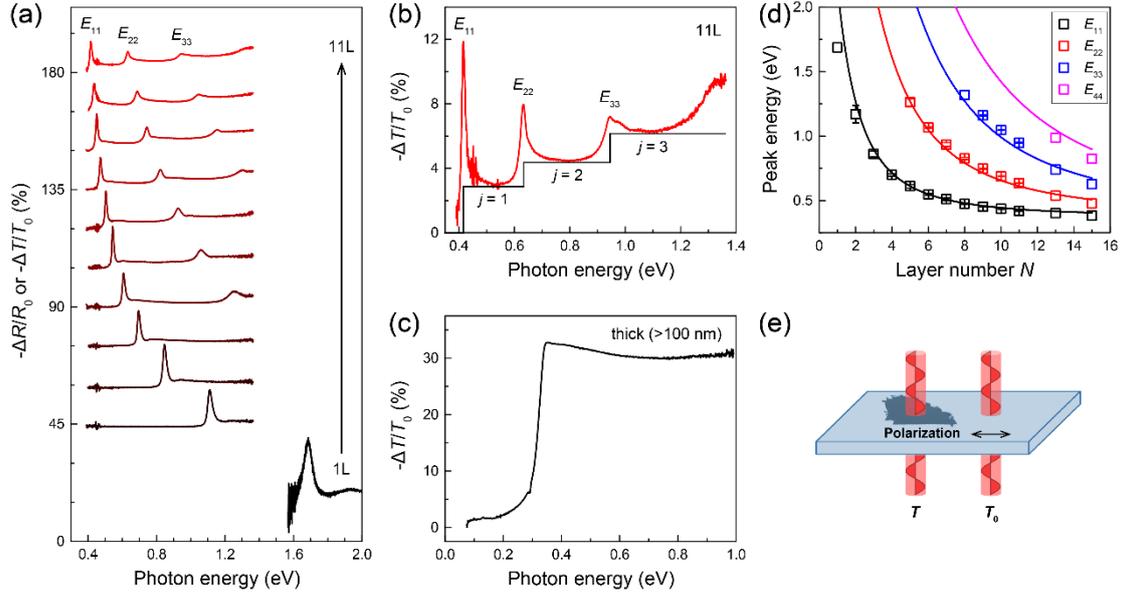

**Figure 6.** Layer-dependent optical absorption in BP. a) Reflectance contrast ($-\Delta R/R_0$) spectrum for 1L BP on SiO$_2$/Si substrate in the visible range, and extinction ($-\Delta T/T_0$) spectra for 2-11L BP on PDMS substrate in the IR range. The polarizations are along the AC direction. b) The extinction spectrum for the 11L BP (the top one in Figure 6a). The black line illustrates the 2D DOS, with step-like features. c) Extinction spectrum for thick BP (thickness > 100 nm) on Si substrate, behaving the same as bulk BP electronically. Reproduced with permission.[42] Copyright 2017, Springer Nature. d) Evolution of the four branches ($E_{11}$, $E_{22}$, $E_{33}$, $E_{44}$) of optical transition energies with layer number, the error bar indicates sample-to-sample variations. The solid lines are global fits to the data using the 1D tight-binding model (Equation (3.1)). Note that the 13L and 15L data are extracted from Figure 7a.[42]

As a representative example, the spectrum for 11L is shown in Figure 6b, three peaks ($E_{11}$, $E_{22}$ and $E_{33}$) can be observed, due to the strong exciton absorption. Above the exciton peaks, a relatively flat "plateau" can be seen, assigned to the continuum absorption due to optical transitions to free-carrier states. The continuum absorption exhibits step-like features, resembling the 2D DOS (illustrated by the black line in Figure 6b). This is a typical character of 2D systems, distinctly different from other dimensionalities (0D, 1D and 3D). As the thickness increases to bulk, as shown in Figure 6c, the excitonic features disappear as expected, with only an absorption edge at ~0.34 eV. Figures 6a and 6c present systematic optical measurements of BP from 1L to bulk, it was found that the optical bandgap ($E_{11}$) sensitively depends on the number of layers, monotonically decreasing from 1.69 eV (1L) to 0.34 eV (bulk), demonstrating a large layer-tunability. Moreover, Zhang et al. revealed that each BP of different thickness exhibits unique absorption features, severing as its optical fingerprints.[42] This provides the most reliable way to determine the layer number of BP.

As summarized in Figure 6d, all branches of optical transitions ($E_{11}$ to $E_{44}$) show significant layer-dependence, with the transition energy decreasing with layer number. This reflects the band structure evolution, and can be well described by a simple 1D tight-binding model.[42,43,121] In this model, a single layer can be viewed as an "atom", then $N$-layer mimics the formation of a molecule by $N$ atoms. When two identical atoms get close to each other, the interatomic interactions split the CB (VB) into two subbands – CB1 and CB2 (VB1 and VB2), one is lifted and the other is lowered with respect to that in the pristine atom. Consequently, the bandgap is reduced in the two-atoms system. For a molecule of $N$ atoms, the CB and VB split into $N$ pairs of subbands, then the bandgap is further reduced.

In the TB model for $N$-layer BP, only two parameters are needed, one is the monolayer bandgap $E_{g0}$, determined by the two intralayer hoppings ($t_\parallel^1$ and $t_\parallel^2$), and the other is the interlayer hopping ($t_\perp$), here denoted as $\gamma_c$ for CB and $\gamma_v$ for VB, respectively. With only the nearest interlayer couplings are taken into consideration, at Γ point of the 2D Brillouin zone ($k_x = k_y = 0$), the energy for optical transition between the $j$ pair of subbands can be expressed as

$$E_{jj}^N = E_{g0} - 2(\gamma_c - \gamma_v)\cos(\frac{j\pi}{N+1}) \tag{3.1}$$

The four branches of optical transitions can be well reproduced by Equation (3.1), the two parameters are extracted to be $E_{g0} = 2.17\ eV$, and $\Delta\gamma = \gamma_c - \gamma_v = 0.90$ eV by the global fitting (solid lines in Figure 6d), showing excellent agreements with the experimental data. With the two parameters in hand, all optical transition energies for each layer number and subband index ($N$, $j$) are available. The results by Zhang et al. build a complete database for optical transitions in BP, and unambiguously uncover the band structure evolution from 2D to 3D,[42] highlighting the important role of interlayer couplings.

Few-layer BP can be viewed as an infinite QW, in which electrons are free to move in the layer (*x-y*) plane, while strictly confined in the out-of-plane (*z*) direction, leading to the quantization of energy bands. For the 13L and 15L cases (Figure 7a), the peak energies of each subband transition $E_{jj}$ can be well described by the QW model $E_{jj} = \frac{\hbar^2\pi^2 j^2}{2\mu_z L_z^2} \propto j^2$, where $L_z$ is the QW width (i.e., the thickness for few-layer BP), and $\mu_z$ is the reduced effective mass in the $z$ direction, defined as $\frac{1}{\mu_z} = \frac{1}{m_{ez}} + \frac{1}{m_{hz}}$. As seen in Figure 7b, the fittings show an excellent agreement with the experimental data. One can find that the energy spacing between adjacent $E_{jj}$ increases with subband index $j$ and decreases with layer number $N$.

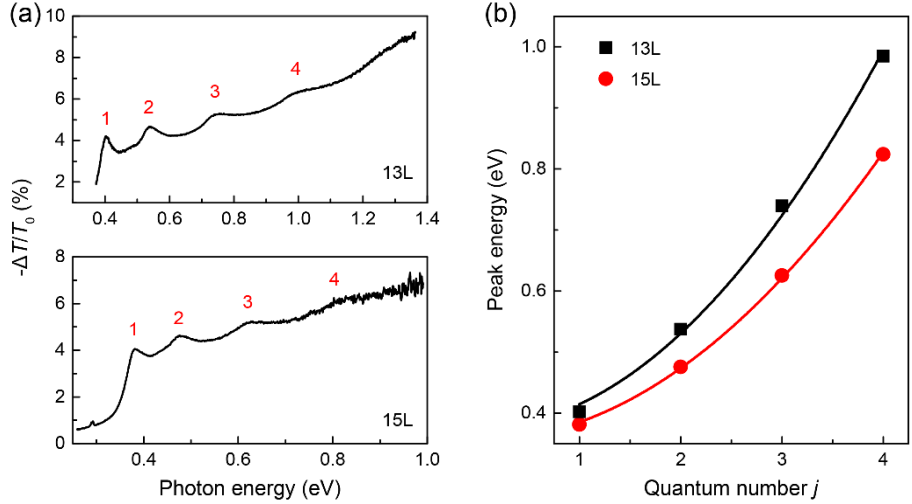

**Figure 7.** Quantum well model. a) IR extinction ($-\Delta T/T_0$) spectra for 13L (upper panel) and 15L (lower panel) BP on quartz substrate. The subband index is indicated in red. Note that these two samples were exfoliated and measured in air, the weak absorption intensity is mainly due to poor sample quality. b) Peak energies as a function of subband index (i.e., quantum number) $j$, the solid lines are fits to the data using the QW model $E_{jj} \propto j^2$. Reproduced with permission.[42] Copyright 2017, Springer Nature.

### 3.4. Forbidden Optical Transitions

Additional features were observed in air-doped few-layer BP.[42] **Figure 8**a shows the IR extinction spectrum of a 9L BP on quartz substrate, exfoliated and measured in air. Other than the main peaks ($E_{11}$ and $E_{22}$), two weak peaks appear in the spectrum between them, indicated by the red arrows. For clarity, the titled background was removed, shown in Figure 8b, the two peaks are well fitted by Lorentz functions. These peaks are assigned to optical transitions between CBs and VBs with different subband index ($\Delta j \neq 0$), i.e., from VB1 to CB2 ($E_{12}$) or VB2 to CB1 ($E_{21}$), also known as hybrid transitions, illustrated in Figure 5c. Similar phenomena were also observed in few-layer BP with other thickness.[42]

Based on the selection rule, in an intrinsic QW, the overlap of wavefunctions between $|j_1\rangle$ and $|j_2\rangle$ subbands is zero, i.e., $\langle j_1|j_2\rangle = 0$ $(j_1 \neq j_2)$,[45,111] and hence hybrid transitions ($\Delta j \neq 0$) are strictly forbidden. However, theoretical studies show that a vertical electric field will relax the selection rule, making hybrid transitions allowed.[111] As shown in Figure 8c, hybrid peaks ($E_{12}/E_{21}$) appear in an electrically doped BP. Moreover, with the increase of doping, the hybrid peaks gradually become dominant, with a large proportion of oscillator strength transferred from the main peaks ($E_{11}/E_{22}$). Experimentally, hybrid transitions have been observed in traditional QWs, e.g., GaAs/AlGaAs.[122,123] For the few-layer BP case (Figures 8a-b), hybrid transitions may be activated by unintentional air doping, which leads to the accumulation of surface charges, giving rise to a vertical electric field. A preliminary work on gate-doped few-layer BP was reported,[124] and further efforts are needed and rewarding, since the hybrid peaks are expected to host important information, e.g., the symmetry between electrons and holes in the $z$ direction.

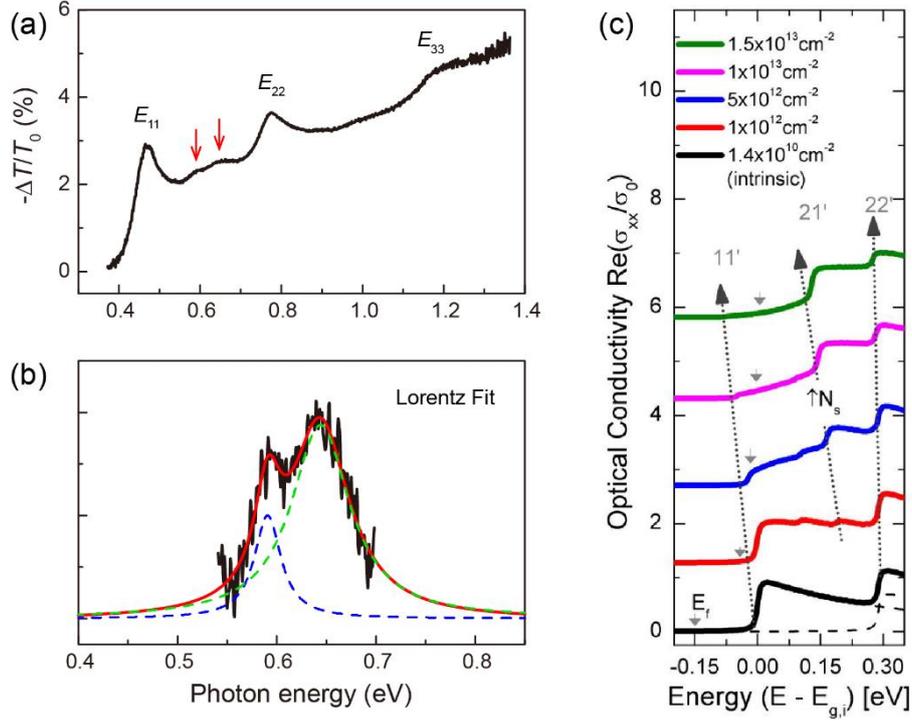

**Figure 8.** Hybrid transitions in few-layer BP. a) IR extinction (-$\Delta T/T_0$) spectrum of air-doped 9L BP on quartz substrate, $E_{12}/E_{21}$ peaks are indicated by the red arrows. b) The $E_{12}/E_{21}$ peaks are extracted from Figure 8a, with the titled background removed for clarity. The Lorentz fits are also shown. Figures 8a-b are Reproduced with permission.[42] Copyright 2017, Springer Nature. c) Calculated optical conductivity spectrum of 5 nm BP, with different electrical doping. Reproduced with permission.[111] Copyright 2016, American Chemical Society.

### 3.5. Optical Signature of BP Degradation

The main drawback of BP is the instability in air, severely limiting its practical applications. Each P atom has five valence electrons, among which only three are engaged in the covalent bonds, leaving behind an unbonded lone pair. This lone pair is very liable to react with $O_2$, leading to the degradation. The entire degradation process contains a two-step chemical reactions:[125] BP first reacts with $O_2$ to form phosphorus oxide (P + $O_2$ → $P_xO_y$), covering its surface, and then $P_xO_y$ further reacts with $H_2O$ to form soluble phosphoric acid ($P_xO_y$ + $H_2O$ → $H_3PO_4$), as a result, a new BP surface is exposed to air. Sequentially, the above chemical process takes place again and again, leading to the continuous degradation of BP. Some techniques have been used to characterize the degradation of BP, such as optical microscope,[126] atomic force microscope (AFM),[125-128] Raman spectroscopy,[125] and electrical measurements.[126,128,129] For these characterizations, the time scale is typically in the day order of magnitude. Besides, the degradation can not be directly linked to the change of the band structure.

In contrast, FTIR serves as a rapid characterization tool, and is able to directly reveal the degradation induced changes in the band structure of BP. **Figure 9**a shows an optical image of a 3L BP on quartz substrate, taken one day after the exfoliation in air. It is clearly seen that a large part of the sample disappeared (indicated by the white dashed line), demonstrating severe degradation in relatively high humidity. Although it is still optically flat for the remaining part, in fact, the sample quality is severely damaged. Wang et al. reported a FTIR study of air-exposed few-layer BP, uncovering the band structure evolution with degradation.[130] In the controlled experiments, the IR extinction spectra of a 6L BP were traced successively in $N_2$ and air, as shown in Figures 9c and 9b, respectively. For the case in air, Figure 9b shows the evolution of the IR extinction spectra with time, revealing that BP degrades very fast in the time scale of minutes. After 10 minutes, a clear change is seen in the spectrum, with a blueshift of the peak position, a reduced peak intensity (not the integrated intensity) and a broadened linewidth both for $E_{11}$ and $E_{22}$. As the time goes on, these changes continue to increase. In particular, the blueshift can act as an optical signature of degradation, which reflects the influence of the quantum confinement on the band structure and can be well understood in the QW model. As the surface is oxidized, the "effective thickness" of BP is reduced, thus the quantum confinement is enhanced, leading to the blueshift. In some cases, $N$-layer BP is right thinned to ($N$-1)-layer by degradation, confirmed by the IR fingerprints. Several groups reported that BP can be thinned layer by layer via heating[131] or laser irradiation,[90] the mechanism is actually heat induced degradation of the upper layers of BP. Wang et al. further revealed that additional defects are introduced in degraded BP, leading to the damage in sample quality, manifested by the increased Stokes shift.[130] This explains the reduced peak intensity and broadened linewidth, since nonradiative decay channels are increased. Besides, the increase of the moisture will significantly accelerate the degradation. And the thinner the BP is, the severe the degradation will be.

On the other hand, it is in great demand to effectively protect BP against degradation. In previous studies, the methods of encapsulation with PMMA,[132] $h$-BN,[133,134] $Al_2O_3$,[126,128] and organic molecules[129,135] are commonly used. FTIR characterization shows that although the degradation slows down, but still takes place in encapsulated few-layer BP in air. Instead, $N_2$ (or other inactive gases) protection works excellently. For the same 6L sample in $N_2$, as shown in Figure 9c, the IR extinction spectrum remains unchanged over time even after one hour, demonstrating perfect sample stability. The results by Wang et al. pave a way for the protection of few-layer BP, largely facilitating the explanation of its intrinsic properties.

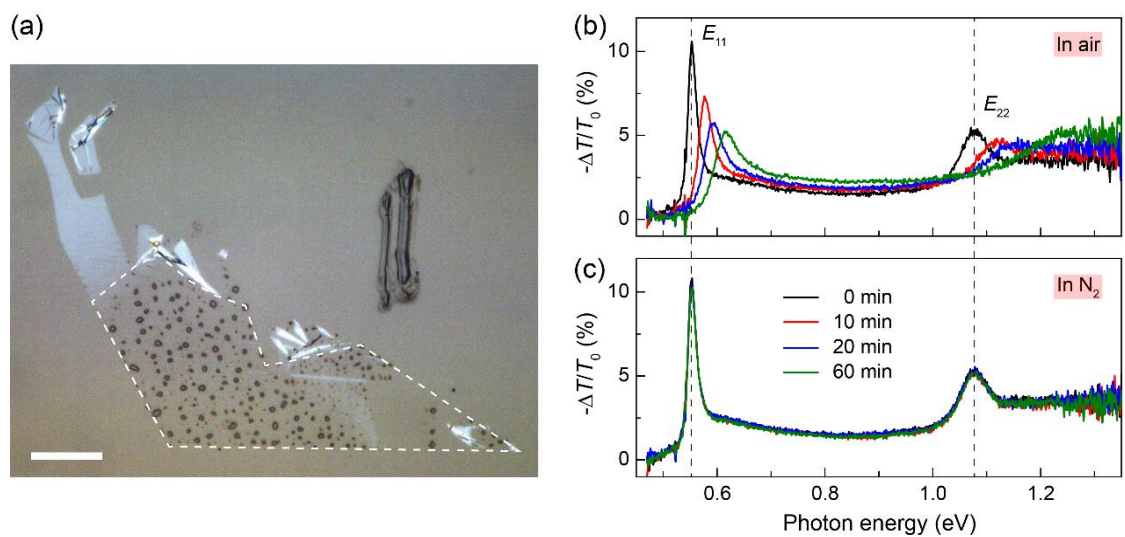

**Figure 9.** Optical characterization of BP degradation. a) Optical image of 3L BP on quartz substrate taken one day after the exfoliation in air, the disappeared part due to degradation is indicated by the white dashed line. The scale bar is 20 μm. IR extinction ($-\Delta T/T_0$) spectra of 6L BP on PDMS substrate b) in air and c) in $N_2$, respectively, measured at different time after the exfoliation. Figures 9b-c are Reproduced with permission.[130] Copyright 2019, American Physical Society.

### 3.6. Layer-Dependent Photoluminescence

PL spectroscopy is directly related to the bandgap transition ($E_{11}$), but is incapable of higher subband transitions. For TMDCs, the bandgap is direct only in monolayer.[25,26] While for BP, the direct-gap character holds from the monolayer to the bulk, suggesting significantly layer-tunable PL. As presented in **Figures 10**a-c, Li et al. reported strong PL in 1-3L BP at 77 K in the visible to near-IR range.[43] It can be see that the peak position for 1L BP is around 1.7 eV, and sharply decreases to 1.1 eV and 0.8 eV in 2L and 3L, respectively. Besides, the PL exhibits strong anisotropy with respect to the detection polarization. Wang et al. further revealed that PL is always linearly polarized in the AC direction, regardless of the excitation polarization.[44]

It is noticed that the PL peak position for 1L BP reported by several groups shows a large discrepancy, ranging from 1.3 to 1.75 eV.[39,42-44,56,136] This may be caused by the following reasons: i) the reported "1L" is possibly not the true 1L, instead a partially degraded 2L. As discussed in the above section, the "effective thickness" is reduced after the degradation, e.g., from 2L to "1.5L", leading to the significant blueshift, with the peak energy higher than that of 2L, but still lower than that of 1L. ii) 1L BP degrades very fast, in this case, defects are easily to be introduced. Therefore, the observed PL is possibly from defects, with the emission energy far below the fundamental bandgap, as has been confirmed by Pei et al.[136]

Recently, Chen et al. reported the first mid-IR PL in few-layer BP based on a FTIR spectrometer.[54] As shown in Figure 10d, the PL exhibits strong layer-dependence, with the peak energy monotonically decreasing with layer number, from 0.441 eV (4.5 nm or ~9L) to 0.308 eV (46 nm or ~92L, bulk) at 80 K. The extracted optical bandgaps are summarized in Figure 10e, well fitted by the 1D tight-binding model (blue line). The two fitting parameters are $E_{g0} = 2.12\ eV$, and $\Delta\gamma = 0.905$ eV, showing excellent agreements with those obtained from optical absorption measurements by Zhang et al.[42] Moreover, Chen et al. found that the (optical) bandgap increases with temperature in bulk BP (46 nm), exhibiting an anomalous behavior.[54] Most strikingly, in a very recent study, Huang et al. found that this anomalous behavior in bulk BP gradually evolves with decreasing thickness, and eventually turns into the normal one in monolayer BP, uncovering the crucial role of interlayer couplings in layered materials.[137] Furthermore, mid-IR electroluminescence (EL) has been realized in bulk BP, taking a step towards the mid-IR luminescent devices.[138]

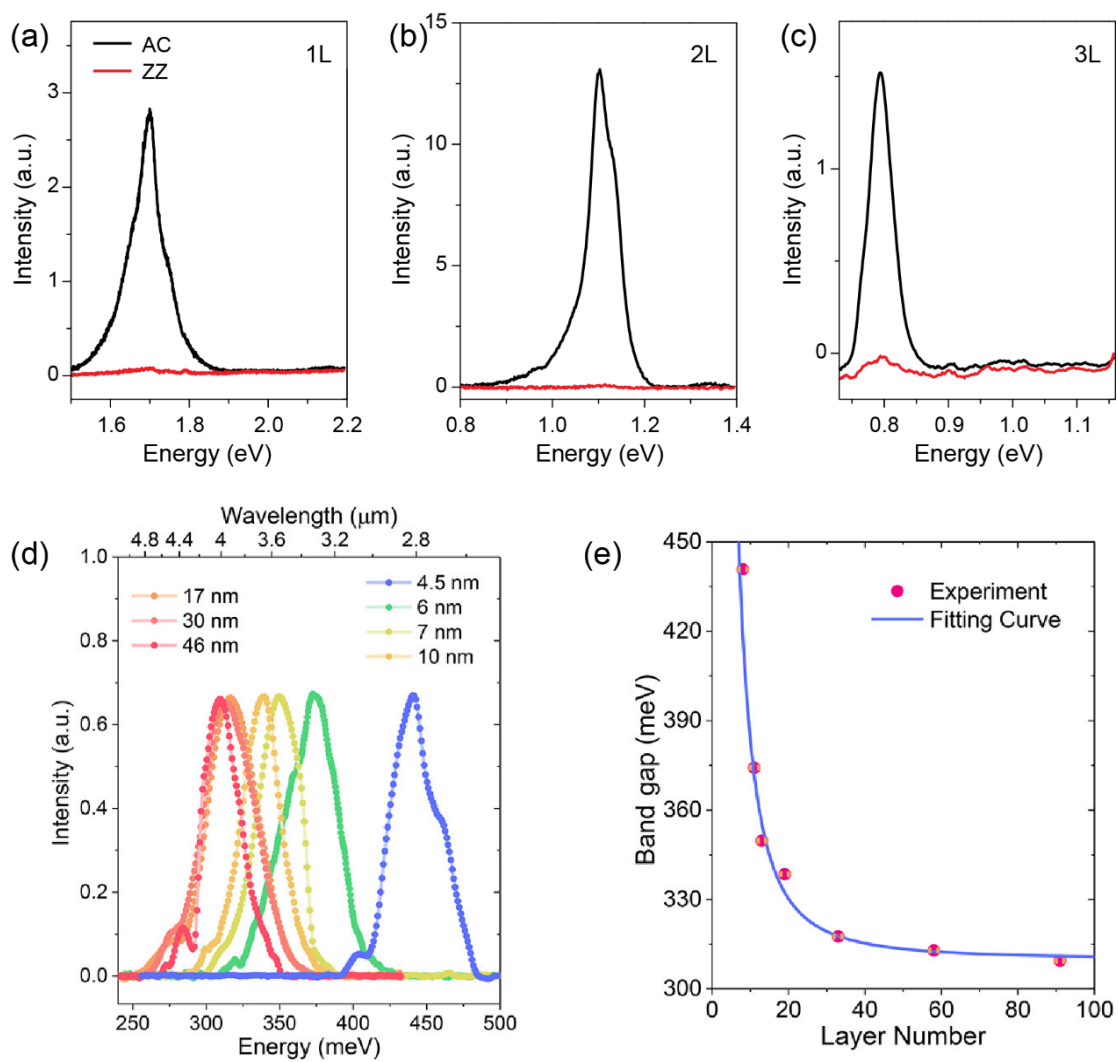

**Figure 10.** Layer-dependent PL in BP. a-c) PL of 1-3L BP under unpolarized excitation at 77 K, and detected in the AC (black line) and ZZ (red line) directions, respectively, in the visible to near-IR range. Reproduced with permission.[43] Copyright 2017, Springer Nature. d) PL of BP with thickness from 4.5 nm (~9L) to 46 nm (~92L, bulk) at 80 K in the mid-IR range, the detection polarization is along the AC direction. e) The optical bandgap, extracted from the PL peaks in Figure 10d, as a function of layer number. The blue line is a fit using the 1D tight-binding model (Equation (3.1)). Figures 10d-e are Reproduced with permission.[54] Copyright 2019, American Chemical Society.

## 4. Band Structure Engineering

2D materials can be viewed as a "surface" and the hosted carriers are mostly exposed to the environment, and hence the physical properties are easily to be tuned by various external techniques. For 2D materials, one of the important topics is to control the band structure to overcome the limit of natural properties. In this section, we will give a summary of the experimental progress in the band structure engineering of few-layer BP, by means of electrical, mechanical and electrochemical methods.

**4.1. Electric-Field Tuning**

Kim et al. reported a widely tunable bandgap in BP by *in-situ* surface doping of K atoms, which denote charges to the upper layers of BP and thus introduce a vertical electric field.[139] The surface bands of BP were monitored by ARPES measurements. With increasing K doping, the bandgap gradually decreases, explained by the well-known Stark effect. At a critical point of the doping, the bandgap is fully closed with a semiconductor-to-semimetal transition. Interestingly, the authors also found that at this critical point, the band dispersion in the AC direction becomes linear, while it remains nearly parabolic in the ZZ direction, resulting in an unusual anisotropic Dirac semimetal state.

In practical applications, electrical gating is a more feasible way. In recent years, significant progress has been achieved in this aspect. In these studies, the bandgap of few-layer BP is extracted from the measurements of electrical transport,[140-142] photoresponse[60] and PL,[61] respectively. Below, we take the last one by Chen et al.[61] as an example, since PL spectroscopy yields a relatively direct link to the bandgap.

**Figure 11**a illustrates the dual-gate device, previously used in bilayer graphene for bandgap opening.[143,144] With the independent top gate and bottom gate, the bandgap and carrier doping concentration can be independently controlled. Figure 11c shows the PL evolution of ~20L BP under different electric fields. With increasing bias voltages from 0 to 40 V (corresponding to an external electric field from 0 to 0.48 V/nm), the PL peak exhibits a continuous redshift from 0.33 to 0.16 eV, indicating a shrink of the bandgap. Their findings are supported by a tight-binding calculation. Figure 11e presents the calculated band structure of ~20L BP near the Γ point, under an electric field of 0 and 0.48 V/nm, respectively. It clearly shows that the electric field shifts the subband CB1 downward and the subband VB1 upward, as a result, the bandgap decreases, but the band shape is barely changed. In contrast, higher subbands are less affected.

Furthermore, it can be seen from Figure 11c that the PL intensity (or oscillator strength) decreases with increasing electric fields. To clarify the origin, the authors calculated the wavefunctions of electrons in CB1 and holes in VB1 at the Γ point, with and without electric fields, respectively. It is found that under an electric field, the electron and hole wavefunctions shift in the opposite directions. Therefore, the overlap between them decreases, leading to the reduction of the PL intensity.

It is worth noted that the tight-binding calculation predicts that with increasing electric fields, higher subband transitions ($E_{22}$, $E_{33}$, …) will also exhibit a redshift, but the shift will be much smaller than that of $E_{11}$. Substantially, PL spectroscopy is only accessible to the bandgap transition ($E_{11}$). By contrast, absorption spectroscopy based on FTIR is of great advantage in characterizations of gate-tunable band structures, though challenging but rewarding in further understanding the electric field effects on 2D subbands. The results by Chen et al. clearly demonstrate that the direct-gap light emission can be widely tuned from 3.3 to 7.7 μm with a moderate electric field of 0.48 V/nm,[61] promising great potentials in BP-based tunable IR devices.

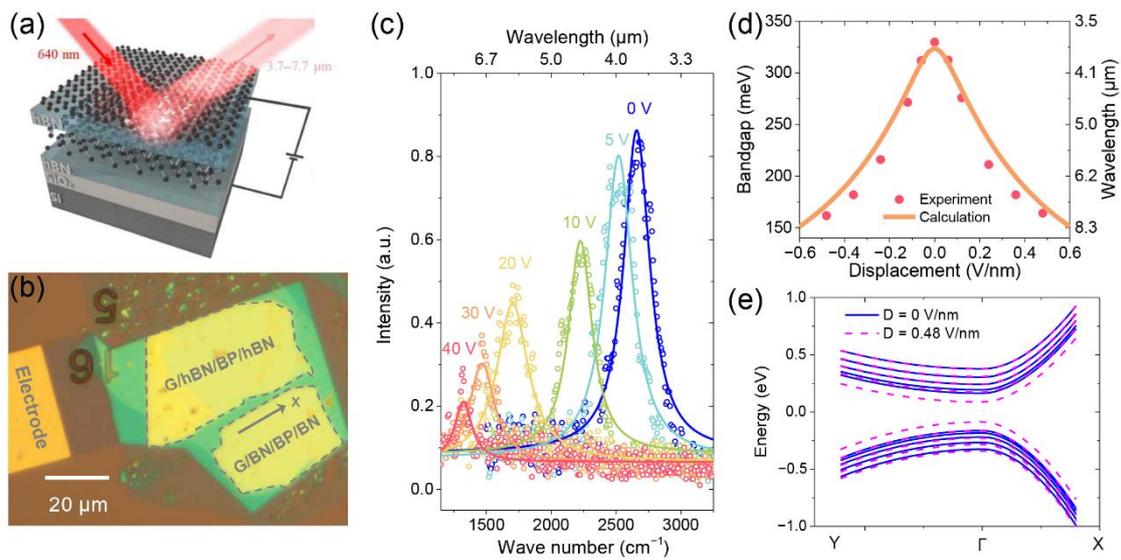

**Figure 11.** Gate-tunable bandgap in BP. a) Schematic of the dual-gated device. The mid-IR PL is measured based on a FTIR spectrometer, with the excitation laser of 640 nm. b) Optical image of the artificially stacked heterostructure: Graphene/BN/BP/BN/SiO$_2$/Si (from top to bottom). c) PL spectra of ~20L BP, under different bias voltages from 0 (intrinsic) to 40 V. d) The (optical) bandgap as a function of the electric field strength, extracted from Figure 11c (dot). The solid line represents the TB calculation. e) Calculated band structure of ~20L BP, under the electric field of 0 and 0.48 V/nm (i.e., 40 V bias), respectively. Reproduced with permission.[61] Copyright 2020, American Association for the Advancement of Science.

## 4.2. Strain Tuning

Strain has been demonstrated as an effective way to tune the band structure of BP both theoretically[73,145-147] and experimentally.[42,62,63,148] Zhang et al. studied the uniaxial strain effect on the band structure of few-layer BP, with a two-point bending method (**Figure 12**a).[42] Figure 12c shows the IR extinction spectra of a 6L BP under different tensile strains up to 0.92%, with the strain applied in the AC (red) and ZZ (blue) directions, respectively. In both cases, the $E_{11}$ and $E_{22}$ peaks exhibit a monotonic blueshift with increasing tensile strain, indicating an increase in $E_g$. As expected, compressive strain induces a decrease in $E_g$ (not shown here). The strain effect on $E_g$ is opposite to the TMDCs case,[149-151] due to the unique puckered structure of BP. As is well-known, BP is an anisotropic material, one may naturally think that the strain effect exhibits a strong direction dependence, in other words, the AC and ZZ strains induce different shift rates of optical transitions. Surprisingly, it is not the case. Figure 12d summarizes the peak positions as a function of tensile strain in both directions, with little difference observed for each strain. From the linearly fittings, it is extracted that the shift rate for $E_{11}$ is 117 meV/% (AC direction) and 124 meV/% (ZZ direction), respectively, and 99 meV/% for $E_{22}$ in both directions. This means that 1% tensile strain can induce a >20% change in the (optical) bandgap of the 6L BP.

The study of bulk BP also demonstrates a nearly isotropic strain effect.[63] This unusual phenomenon can be well understood within the framework of the tight-binding model. Although in-plane strain can also affect the interlayer coupling, but mostly affects the intralayer bonding. In other words, for $E_{jj}$ in Equation (3.1), in-plane train has a strong effect on $E_{g0}$ and a smaller effect on $\Delta\gamma$. For simplicity, the monolayer is taken as an example in the following discussion. The monolayer bandgap can be expressed as $E_{g0} = 2(2t_\parallel^1 + t_\parallel^2)$, with $t_\parallel^1 < 0$, $t_\parallel^2 > 0$.[85] Under strain, the change of $E_{g0}$ can be quantitatively expressed as (in the unit of eV)[152-154]

$$\Delta E_{g0} = 4.1\varepsilon_x + 5.7\varepsilon_y - 12.9\varepsilon_z \qquad (4.1)$$

where $\varepsilon_x$, $\varepsilon_y$, and $\varepsilon_z$ denote the uniaxial strain along the $x$ (AC), $y$ (ZZ), and $z$ (out-of-plane) direction, respectively. Note that monolayer BP consists of two sublayers (illustrated in Figure 12b), $\varepsilon_z$ corresponds to the inter-sublayer distance ($d$). According to Equation (4.1), it first looks like that the ZZ strain is more effective than the AC strain in terms of the bandgap tuning. But in fact, $\varepsilon_z$ will contribute to the in-plane strain effect as well. DFT calculations predict a positive Poisson's ratio in the AC direction ($v_{zx} > 0$), but a negative one in the ZZ direction ($v_{zy} < 0$), this leads to the opposite sign of $\varepsilon_z$.[50] In other words, tensile strain in the AC direction will reduce $d$, inducing an additional compressive strain in the $z$ direction ($\varepsilon_z < 0$). While for the ZZ case, $\varepsilon_z > 0$. The cancelation from $\varepsilon_z$ leads to the nearly isotropic strain effect. According to Equation (4.1), when ($v_{zx}$ - $v_{zy}$) = 0.124, the cancelation is exact, in reasonable agreement with the DFT calculations (0.073).[50] With $v_{zx}$ = 0.046 ($v_{zx}$ = -0.027), the deduced shift rate for the monolayer is 46.9 meV/% (53.5 meV/%), smaller than the experimentally obtained value for the 6L BP. This indicates that the in-plane strain effect on the interlayer coupling ($\Delta\gamma$) is non-negligible, suggesting strong layer-dependence of the strain effect, which will be discussed in the following section.

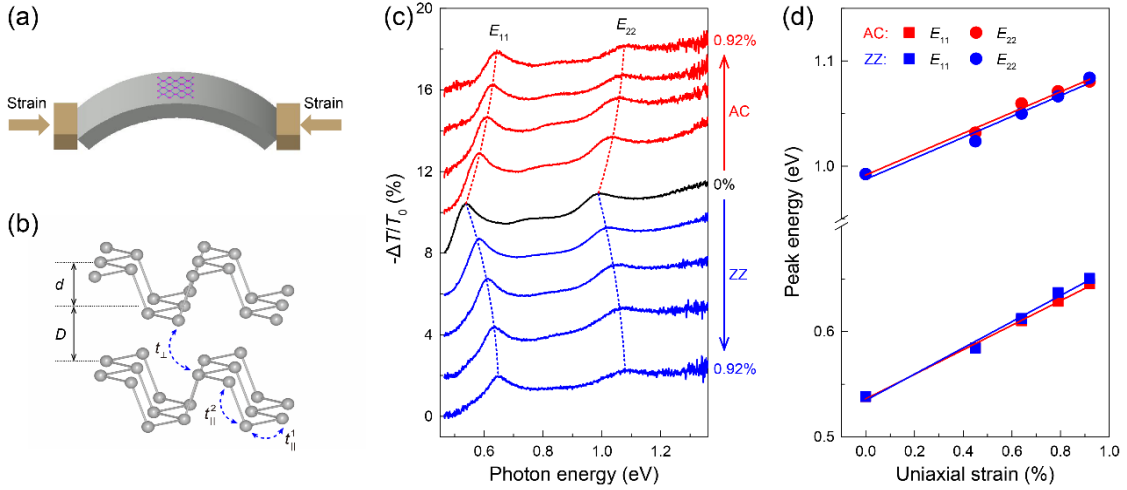

**Figure 12.** Band structure engineering by uniaxial strain in few-layer BP. a) Schematic of the uniaxial strain setup. b) Lattice structure of bilayer BP, with two intralayer hoppings ($t_\parallel^1$ and $t_\parallel^2$) and one interlayer hopping ($t_\perp$) indicated by blue arrows. c) IR extinction ($-\Delta T/T_0$) spectra of 6L BP under varying uniaxial strains, with the strain along the AC (red) and ZZ (blue) directions, respectively. The dashed lines trace the evolution of the $E_{11}$ and $E_{22}$ peak positions with uniaxial strain. d) Peak energies of $E_{11}$ and $E_{22}$ of the 6L BP, as a function of the uniaxial strain. The solid lines are linear fits to the data. Reproduced with permission.[42] Copyright 2017, Springer Nature.

Biaxial strain is expected to be more effective than uniaxial strain. Huang et al. systematically studied the biaxial strain effect on the band structure in 2-10L BP.[62] **Figure 13**a illustrates the experimental setup, the substrate expands (contracts) when heated (cooled), then biaxial tensile (compressive) strain will be introduced to few-layer BP. In this work, polypropylene (PP) is selected as the substrate, for the reason that it exhibits a large thermal expansion coefficient but a small heat conductivity, thus the temperature effect on BP can be ignored. Figure 13c shows the IR extinction spectra of a 6L BP, under different biaxial tensile ($\varepsilon > 0$) and compressive ($\varepsilon < 0$) strains. Similar to the uniaxial strain case (Figure 12c), the $E_{11}$ and $E_{22}$ peaks both exhibit a monotonic blueshift with increasing strain, but at a larger shift rate of 222 and 167 meV/%, respectively. These values are roughly twice of those in the uniaxial strain case, consistent with expectations. Most importantly, $E_{11}$ shifts faster than $E_{22}$, in other words, the energy separation between $E_{11}$ and $E_{22}$ ($\Delta E_{21} = E_{22} - E_{11}$) decreases with increasing strain, similar phenomenon has also been observed in the uniaxial strain case. In the tight-binding model, $\Delta E_{21}$ acts as a gauge of the interlayer coupling ($\Delta \gamma$) strength. As shown in the inset of Figure 13d, $\Delta E_{21}$ inversely scales with strain at a linear slope of -55 meV/%, indicating a weakened (enhanced) interlayer coupling with tensile (compressive) strain, in sharp contrast to the common sense. Further studies of BP with different thickness (2-10L) reveal that the strain effect is strongly layer-dependent, as summarized in Figure 13b. This can be well understood in the tight-binding model.

As seen from Equation (3.1), the transition energy $E_{jj}$ in few-layer BP is determined by two parts, i.e., the monolayer bandgap $E_{g0}$ and the interlayer coupling term $2\Delta\gamma \cos(\frac{j\pi}{N+1})$. It is reasonable to assume that $E_{g0}$ and $\Delta\gamma$ both change linearly with strain as follows,

$$E_{g0} = E_{g0}^0 + h \cdot \varepsilon, \ \Delta\gamma = \Delta\gamma_0 + k \cdot \varepsilon \qquad (4.2)$$

where *h* and *k* are fitting parameters, describing the change rates of the monolayer bandgap and the interlayer coupling strength under strain, respectively. With Equations (3.1) and (4.2), the strain induced shift rate is deduced as

$$\frac{dE_{jj}}{d\varepsilon} = h - 2k\cos(\frac{j\pi}{N+1}) \qquad (4.3)$$

According to Equation (4.3), it concludes that the shift rate of $E_{jj}$ is layer (*N*) dependent and transition order (*j*) dependent for $k \neq 0$. The global fitting of $E_{11}$, $E_{22}$, and $E_{33}$ using Equation (4.3) is shown in Figure 13b (solid lines). The overall trend of the fitting is in reasonable agreement with the experiment results. Besides, *h* and *k* are extracted to be 66 and -86 meV/%, respectively. With *k* = -86 meV/% and $\Delta\gamma$ = 0.90 *eV*, it straightforwardly indicates that 1% in-plane biaxial strain can induce a ~10% change of the interlayer coupling strength ($\Delta\gamma$).

It should be emphasized that *k* is negative, this means that the interlayer coupling is weakened under tensile strain, in sharp contrary to the common sense. This astonishing phenomenon originates from the puckered structure of BP. The findings by Huang et al. are supported by DFT calculations,[62] under 1% biaxial tensile strain, the interlayer distance (*d* + *D*) decreases by 0.031 Å. However, at the same time, the distance between two sub-layers (*d*) decreases even more by 0.087 Å. The overall result is that the "effective" distance (*D*) between two adjacent BP layers increases by 0.056 Å, leading to the weakened interlayer coupling and enlarged bandgap.

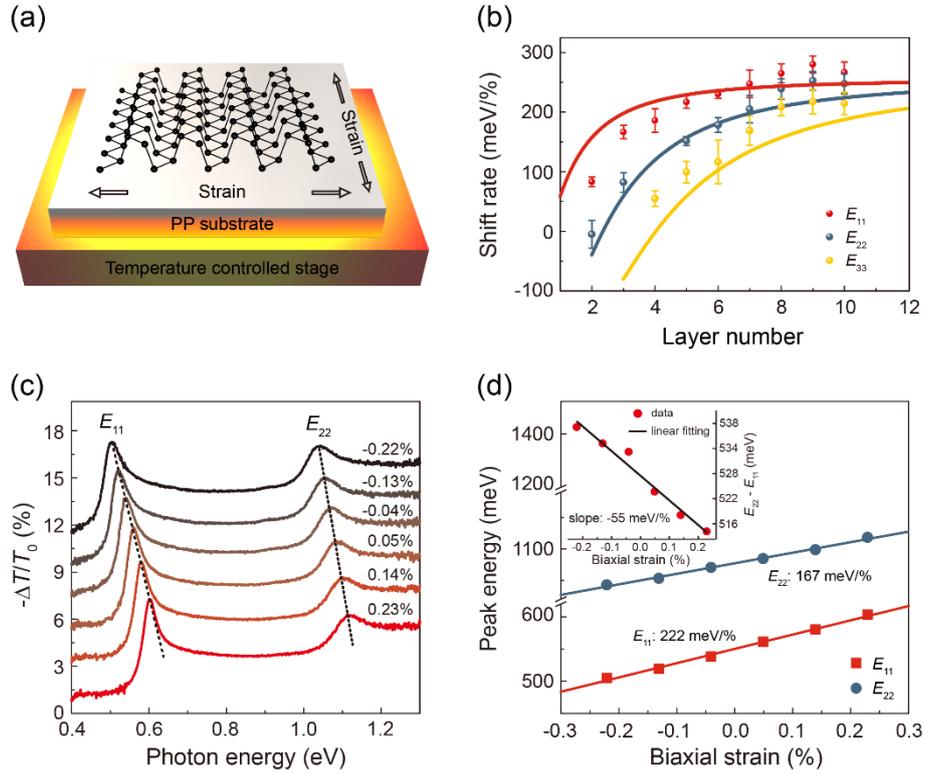

**Figure 13.** Band structure engineering by biaxial strain in few-layer BP. a) Schematic of the biaxial strain setup. b) Biaxial strain induced shift rate as a function of layer number, for $E_{11}$, $E_{22}$, and $E_{33}$ transitions in 2-10L BP. The solid curves are fitted to the data using the tight-binding model (Equation (4.3)). The error bar indicates sample-to-sample variations. c) IR extinction ($-\Delta T/T_0$) spectra of 6L BP, under varying tensile ($\varepsilon > 0$) and compressive ($\varepsilon < 0$) strains. For clarity, the spectra are vertically offset. The dashed lines trace the evolution of the $E_{11}$ and $E_{22}$ peak positions with biaxial strain. d) Peak energies of $E_{11}$ and $E_{22}$ of the 6L BP, as a function of biaxial strain. The solid lines are liner fits to the data. Inset: ($E_{22} - E_{11}$) as a function of biaxial strain, with a linear fit. Reproduced with permission.[62] Copyright 2019, Springer Nature.

### 4.3. Molecular Intercalation

Both in fundamental research and device applications, there is a long-standing desire for air-stable and large-area 2D BP. A recent work by Wang et al. may emerge as a promising and feasible solution to this issue. They reported that a bulk BP crystal can be turned into a stack of monolayers by an electrochemical molecular intercalation method, with the merits of reasonable lateral size and stability maintained (illustrated in **Figure 14**a).[155] In layered materials, the physical properties are largely governed by interlayer couplings. One can imagine that if the interlayer distance in a bulk material is large enough, the interlayer couplings will be negligibly weak, and hence the bulk will behave similarly as the monolayer. One natural example is ReS$_2$, with a large interlayer distance of 6.0-6.9 Å, it has been demonstrated that bulk ReS$_2$ is electronically and vibrationally decoupled.[156] Based on this idea, the authors make use of large-size organic molecules (cetyltrimethylammonium bromide, CTAB), instead of the commonly used lithium ions,[157] to enlarge the interlayer distance of bulk BP.

Figures 14b-f show the structural, optical, and electrical characterizations of the stack of BP monolayers. As indicated by Figures 14c and 14d, the electrochemical intercalation process undergoes several stages: from bulk to few-layer, and finally to monolayer. Note that the resulting "monolayer" is not an individual one, but a stack of monolayers, termed as monolayer phosphorene molecular superlattices (MPMS). PL mapping shows high uniformity of the MPMS flake (Figure 14e). As shown in Figure 14b, transmission electron microscopy (TEM) characterization clearly reveals that the interlayer distance largely increases from 5.24 Å in the pristine bulk BP to 11.21 Å in the resulting MPMS. Furthermore, the authors checked the stability of the MPMS. As seen in Figure 14f, the MPMS devices show little electrical degradation even after 300 h exposure in ambient conditions. In contrast, the pristine BP devices show a sharp drop in the electrical currents after 20-30 h exposure. This demonstrates greatly improved stability of MPMS. The results by Wang et al.[155] push a big step forward towards 2D BP-based device applications.

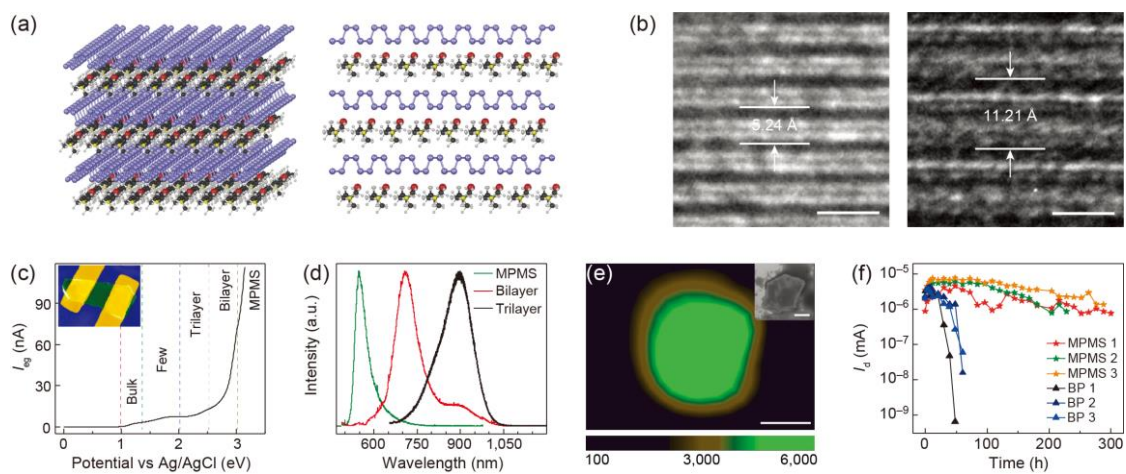

**Figure 14.** Molecular intercalation of BP. a) Schematic illustration of the atomic structure of the MPMS, with BP monolayers separated by CTAB molecules. b) TEM characterization of the interlayer distance in the pristine bulk BP (left) and the resulting MPMS (right), respectively. c) The recorded electrochemical currents as a function the electrochemical potential. d) PL spectra of the intercalated trilayer (green), bilayer (red), and monolayer (termed as MPMS, black) BP, respectively. e) PL mapping at 550 nm of the MPMS. Inset: the corresponding scanning electron microscope (SEM) image. Scale bars: 3 μm. f) Drain current as a function of the exposure time, in comparison of three MPMS (stars) and three pristine BP (triangles) devices. Reproduced with permission.[155] Copyright 2018, Springer Nature.

## 5. Layer-Dependent Excitons

An exciton is a bound electron-hole pair, a two-particle system resembling the hydrogen atom. In the above discussions, the optical transitions are interpreted in the single-particle picture, without excitonic effects involved. But in fact, strongly bound excitons are formed in low-dimensional semiconductors, due to enhanced quantum confinement and reduced dielectric screening of electron-hole interactions, which dominate the optical properties even at room temperature. Excitons can significantly enhance the light-matter interactions, for instance, the optical absorption in a single layer of TMDCs is surprisingly large (>10%) at exciton resonances.[158,159] By contrast, it is only 2.3% in graphene, in which excitons are not involved.[107,109]

The exciton binding energy ($E_b$), describing the energy needed to separate the electron and hole apart, acts as a gauge of the strength of Coulomb interactions. In bulk materials, such as GaN and GaAs, the binding energy is small (typically several to tens of meV)[160]. While in 2D materials, it is significantly enhanced, for instance, $E_b$ is found to be as large as ~500 meV in monolayer TMDCs.[161] From 3D to 2D, the quantum confinement only contributes a fourfold enhancement, most of the rest arises from the reduced dielectric screening. For 2D excitons, the latter plays a dominant role in determining the binding energy.[162] Moreover, it is found that the dielectric screening effect increases with thickness in 2D materials,[163] suggesting significantly layer-dependent excitonic properties.

BP is a layered material with strong interlayer couplings and intrinsic in-plane band anisotropy, most of all, the bandgap is always direct for each thickness. These characters establish BP the ideal platform to study the layer-dependence of excitonic properties, which is inherently challenging for TMDCs because the bandgap is only direct for the monolayers. Exciton physics is a very important topic for 2D materials. In recent years, much attention has been paid to the study of excitons in monolayer TMDCs, including measurements of exciton binding energy,[114-117] exciton diffusion,[164,165] dark exciton states,[166,167] trions[168,169] and biexcitons,[170] electric[171] and magnetic[172,173] fields tuning, exciton-cavity couplings,[174,175] et al. In contrast, related studies on 2D BP is rather limited. In this section, we will present the experimental progress of excitonic properties in 2D BP, highlighting the strong layer-dependence character.

## 5.1. Exciton Binding Energy

Tran et al. found that the exciton wavefunction in monolayer BP is of anisotropic spatial distribution, highly extended along the AC direction. This is due to the anisotropic band dispersion, the carriers are less mobile in the ZZ direction, and thus are more strongly bound. Besides, they also found that the exciton wavefunction in trilayer BP spreads to "each layer" in the *z* direction, thus each of the three layers will contribute to the dielectric screening of electron-hole interactions.[72,119] Qiu et al. point out that although the effective mass in the ZZ direction is much larger than that in the AC direction for monolayer BP, but it is still far from the 1D limit as the effective mass along one direction goes to infinity, thus monolayer BP still exhibits a 2D quantum confinement with the DOS remaining a step function.[176] Several groups calculated that the exciton binding energy in free-standing monolayer BP is as large as ~800 meV.[58,59,72,177-179] Rodin et al. shows that $E_b$ decreases as the substrate dielectric constant increases. For instance, in free-standing monolayer BP, $E_b$ is calculated to be 760 meV, with $SiO_2$ supporting, it sharply reduces to ~400 meV.[177] Moreover, it is also theoretically found that $E_b$ is strongly layer-dependent in 2D BP.[58,59,178]

In experiments, there are two kinds of methods to determine the exciton binding energy. The most straightforward one is to directly measure the ground state (1*s*) energy and the quasiparticle bandgap, commonly using combined techniques of optical spectroscopy and scanning tunneling spectroscopy (STS).[117] An alternative but the most used one is to measure the ground state (1*s*) and excited state (one-photon allowed *s*-series or two-photon allowed *p*-series) energies, and then to indirectly deduce $E_b$.[114-116] Generally, the ground state exciton dominates the optical spectrum, and the feature of excited state excitons is rather weak. In the 2D hydrogen model, the oscillator strength scales with $1/(n-1/2)^3$ ($n$ = 1, 2, 3, …), this indicates that the oscillator strength of 2*s* state is only 1/27 of that of 1*s* state, and it is even smaller for 3*s*, 4*s*, … Therefore, the weak feature of excited state excitons depends sensitively on the sample quality, and the determination of $E_b$ is rather challenging.

Recently, Zhang et al. reported the observation of 2*s* states in 2-6L BP on PDMS substrates at room temperature using FTIR-based absorption spectroscopy.[57] The exfoliated few-layer BP on PDMS shows impressively high sample quality, one typical example (4L) is shown in **Figure 15**a, a sharp and strong 1*s* peak can be seen. On its high-energy side, a weak 2*s* peak is well-resolved, followed by a step-like continuum absorption, resembling the 2D DOS. As summarized in Figure 15b, the 1*s*-2*s* separation is ($\Delta_{12}$) extracted for 2-6L BP (dots), showing excellent agreements with theoretical calculations (diamonds). Based on the 2D hydrogen model, $E_b$ will be directly deduced with the obtained $\Delta_{12}$, as $E_b = 9/8\Delta_{12}$. However, it is found that this relation breaks down in atomically thin 2D materials, such as monolayer TMDCs, due to the nonlocal dielectric screening of Coulomb interactions.[114,115] In this case, $\Delta_{12}$ accounts nearly 1/2 of $E_b$, this can be used to roughly estimate the value of $E_b$. With the parameters to calculate $\Delta_{12}$ in Figure 15b, $E_b$ for 2-6L BP are further obtained by theoretical calculations, summarized in Figure 15c (red solid dots). As expected, $E_b$ shows strong layer-dependence.

Olsen et al. proposed a modified Hydrogen model for nonlocally screened 2D excitons, in which $E_b$ is believed to be determined by the 2D polarizability ($\alpha_{2D}$) and does not directly depend on the effective mass, in sharp contrast to the 3D case.[162] $\alpha_{2D}$ is strongly confined in the 2D plane.[163] Note that $\alpha_{2D}$ has a unit of length (Å), and relates to the in-plane screening length as $r_0 = 2\pi\, \alpha_{2D}$ in vacuum ($\varepsilon_0 = 1$).[180] The physical nature is that $r_0$ (or $\alpha_{2D}$) represents the characteristic length of 2D excitons: at short range (the electron-hole distance $r < r_0$), electrons and holes are weakly interacted with a screened Keldysh potential, while at long range ($r > r_0$), it reduces to the well-known Coulomb potential.[180,181] In the modified Hydrogen model by Olsen et al., $E_b$ can be simply expressed as $E_b = \frac{3}{4\pi \alpha_{2D}}$ for free-standing cases (in atomic units), if the 2D polarizability is large, and this simple relation is found to apply well to kinds of 2D TMDCs.[162] Besides, Tian et al. found that $\alpha_{2D}$ scales linearly with thickness in 2D materials,[163] this suggests a $1/N$ dependence of $E_b$. However, in a real situation, 2D materials are commonly supported on a substrate, which will contribute additional screening (illustrated in the inset of Figure 15c). In such case, Zhang et al. proposed an effective polarizability $\alpha_{\text{eff}}(N) = \alpha_0 + N\alpha_1$, where $\alpha_0$ and $N\alpha_1$ account for the screening from the underlying substrate and the 2D material itself, respectively. Then, $E_b$ is expressed as[57]

$$E_b = \frac{3}{4\pi(\alpha_0 + N\alpha_1)} \tag{5.1}$$

Equation (5.1) excellently explains the layer-dependence of $E_b$ (red line in Figure 15c), and the substrate screening induced reduction in $E_b$ as well. The two extracted parameters are $\alpha_0 = 6.5$ Å and $\alpha_1 = 4.5$ Å, respectively. The value for $\alpha_1$ matches well with theoretical calculations (4.1 Å),[177] indicating that the model (Equation (5.1)) is reasonable. Furthermore, with $\alpha_0 = 0$ and $N = 1$, $E_b$ for free-standing monolayer BP is inferred to be 762 meV, in good agreement with ab initio calculations within the GW-BSE approach (740-800 meV[72,178]). For monolayer BP on PDMS, $E_b$ is significantly reduced to 316 meV, demonstrating the important role of the substrate screening.

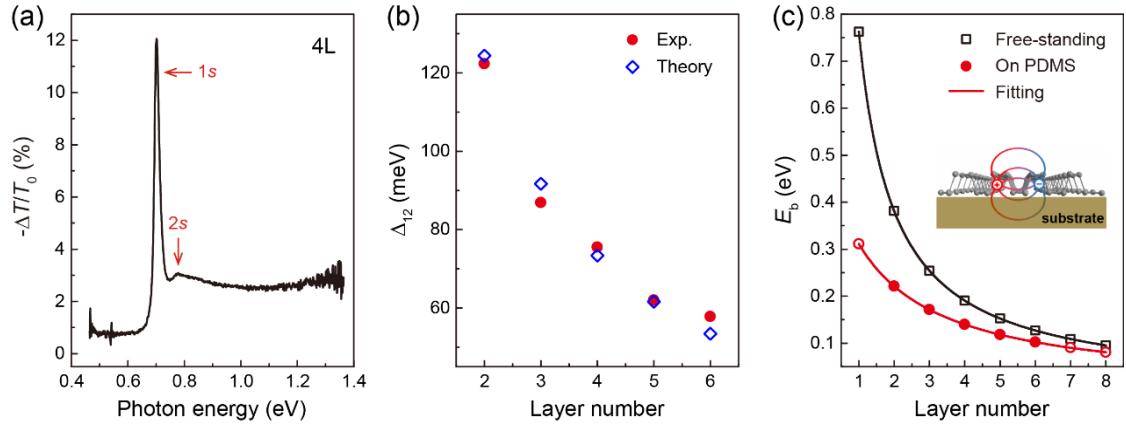

**Figure 15.** Exciton binding energy in few-layer BP. a) IR extinction ($-\Delta T/T_0$) spectrum of 4L BP on PDMS substrate, the two arrows indicate the ground state (1$s$) exciton and the excited state (2$s$) excitons. b) The experimentally and theoretically obtained 1$s$-2$s$ separation ($\Delta_{12}$) as a function of layer number $N$. c) Exciton binding energy $E_b$ of free-standing and PDMS-supported 2D BP as a function of layer number $N$, respectively. The red solid line is a fit to the data using Equation (5.1), and the blank one represents the 1/$N$ scaling. Reproduced with permission.[57] Copyright 2018, American Association for the Advancement of Science.

### 5.2. Exciton and Continuum Absorption Strength

Further, the layer-dependence of the absorption strength is examined in few-layer BP by Zhang et al. With high-quality samples and careful absorption measurements using FTIR, the optical conductivity $\sigma(\hbar\omega)$ was determined for 2-7L BP for the first time.[182] **Figure 16**a presents a typical $\sigma(\hbar\omega)$ spectrum (6L), from which the integrated conductivity (or absorption) of exciton peaks (shaded areas) can be obtained, directly related to the oscillator strength.[183,184] The extracted values for $E_{11}$ and $E_{22}$ excitons in 2-7L BP are summarized in Figure 16b, showing a strong layer-dependence. Surprisingly, the exciton absorption decreases with increasing thickness, in dramatic contrast to 2D TMDCs.[25,158] This interesting phenomenon can be qualitatively understood in the QW model with a $1/N$ relation,[183,184] which means stronger absorption for thinner BP, but predicts a much stepper decrease than the observed (black dashed line in Figure 16b). The deviation arises from the additional dielectric screening from the PDMS substrate. Taking the substrate screening effect into consideration, the relation is corrected to be $1/(\alpha_0 + N\alpha_1)$, with $\alpha_0 = 6.5$ Å and $\alpha_1 = 4.5$ Å defined in the above section, which shows a reasonable fitting (red line in Figure 16b).

In addition, as indicated by the arrows in Figure 16a, the continuum absorption close to the band edge ($\hbar\omega \approx E_g$) of 2-7L BP are summarized in Figure 16c, showing an universal behavior, almost independent of the thickness. The finding is reasonably explained by the authors with an analytical expression for the continuum absorption of 2D materials,

$$A(\hbar\omega) = \frac{g_s g_v \pi \alpha}{2} \cdot \frac{E_g}{\hbar\omega} \cdot \sqrt{\frac{\mu_y}{\mu_x}} \cdot \Theta(\hbar\omega - E_g) \quad (5.2)$$

where $\alpha$ is the fine structure constant (1/137), $\mu_{x(y)}$ is the reduced effective mass in the AC (ZZ) direction, and $g_s$ and $g_v$ denote the spin and valley degeneracy, respectively. $\Theta(\hbar\omega - E_g)$ is a step function, describing the 2D DOS. With $\hbar\omega \approx E_g$, and $g_s = 2$, $g_v = 1$ for few-layer BP,[45] the absorption (Equation (5.2)) is deduced to be $\pi\alpha\sqrt{\frac{\mu_y}{\mu_x}}$, strongly related to the universal value of graphene ($\pi\alpha$), with a prefactor due to the band anisotropy. Furthermore, the authors draw a generalized picture for the continuum absorption in 2D materials: it is not determined by the number of layers, but the number of subbands involved in the optical transitions. As in graphene, the absorption is strictly a constant ($\pi\alpha$),[185] in $N$-layer graphene, it increases to $N\pi\alpha$, since $N$ pairs of subbands are involved. While for $N$-layer BP, only one pair of subbands is involved for the first subband transition ($T_{11}$), and thus only an quanta $\pi\alpha$ is contributed, when the second subband transition ($T_{22}$) occurs, another $\pi\alpha$ is counted.

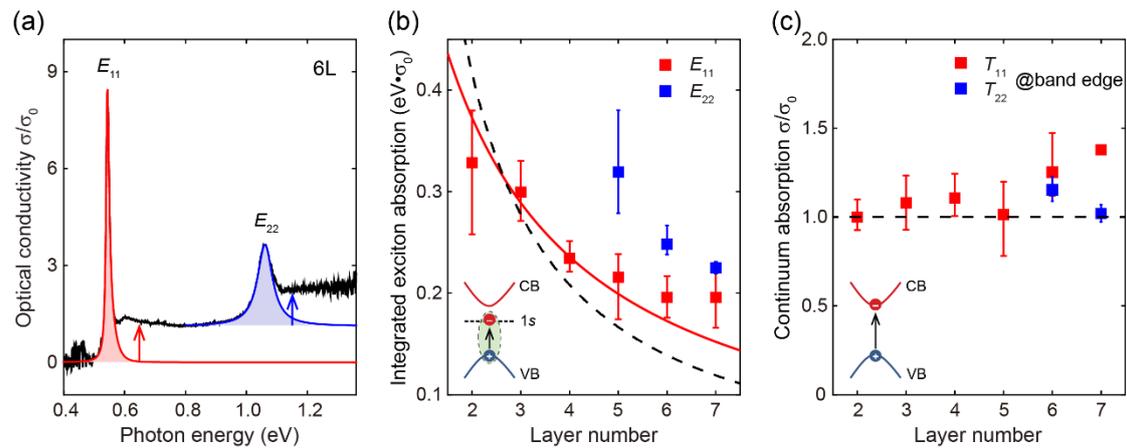

**Figure 16.** Exciton and continuum absorption in few-layer BP. a) Optical conductivity $\sigma(\hbar\omega)$ spectrum of 6L BP, in the unit of $\sigma_0 = e^2/4\hbar$. The fittings of $E_{11}$ and $E_{22}$ exciton peaks are shown. b) The integrated conductivity of excitons as a function of layer number $N$, in the unit of eV $\sigma_0$. The solid curve is the fitting to the $E_{11}$ data using the relation $1/(\alpha_0 + N\alpha_1)$. The dashed curve shows the $1/N$ relation for guideline. c) The continuum absorption ($T_{11}$ and $T_{22}$) close to the corresponding band edge ($\hbar\omega \approx E_g$) is plotted as a function of layer number $N$. The black dashed line indicates the well-known universal conductivity of graphene ($\sigma_0$) for guideline. The error bar indicates sample-to-sample variations. Reproduced with permission.[182] Copyright 2020, Springer Nature.

## 6. Summary and Outlook

In this review, we focus on 2D BP (monolayer and few-layer), an emerging 2D material with intrinsic in-plane anisotropy and highly tunable direct bandgaps. We present a comprehensive summary of the electronic band structures, optical absorption and PL, the band structure engineering, and excitonic properties, with a highlight on the strong layer-dependence characters. BP exhibits unusually strong interlayer couplings due to the unique puckered structure, which endows it the significantly layer-tunable properties. Along with the moderate bandgap and relatively high carrier mobility, BP demonstrates great potentials in flexible and tunable IR devices. This review is expected to stimulate increasing interest in 2D BP, both in fundamental research and device applications. Although, significant progress has been achieved in the fundamental properties of 2D BP. There are still research fields far to be reached or just emerging. Below, we will give a brief outlook.

### 6.1. Twisted BP Heterostructures

Twisted heterostructures (or moiré heterostructures), made of artificially stacked 2D layers with a twist angle, provide a new strategy for modifying the electronic structures by the moiré potential. Many exotic quantum phenomena have been observed in this new material systems, mainly based on isotropic 2D materials, such as graphene[31,32] and TDMCs.[33-36] In these heterostructures, the twist angle is small (typically less than a few degrees). Very recently, Zhao et al. reported a new-type of moiré heterostructures, composed of anisotropic 1L/2L BP with a relatively large twist angle (19°).[186] They observed a new optical resonance in the heterostructures, far below the fundamental transitions of the individual 1L and 2L BP. Besides, this new optical resonance exhibits strong anisotropy, with the principle axis differing from that in any of the two components. The results by Zhao et al. highlight the exciting opportunities to explore moiré physics in anisotropic 2D materials.

Given that BP has a strong layer-tunable direct bandgap and host a series of subband transitions, there exist a huge combinations of BP heterostructures, such as 4L/5L, 3L/9L, et al. Moreover, beyond the bandgap transition ($E_{11}$), it is still not yet reached for the higher subband transitions ($E_{22}$, $E_{33}$, …). Few-layer BP provides an unique opportunity to probe the influence of the twist angle on higher subband transitions. Fascinatingly new physics is expected in such material systems.

**6.2. Higher-order Excitonic States**

In 2D semiconductors, the optical properties are dominated by excitons. It is very essential to deeply understand the exciton physics, especially in 2D materials with intrinsic band anisotropy. Experimentally, only the ground 1$s$ state and a low-lying excited 2$s$ state were observed in few-layer BP.[57] It is far to be reached for other excited states, especially for the $p$-series. Chaves et al. theoretically predicts that the effective mass anisotropy lifts the degeneracy of the 2$p_x$ and 2$p_y$ states, consequently, the 2$s$ state becomes higher than the 2$p_x$ state, while lower than the 2$p_y$ state in few-layer BP,[58] this is dramatically different from the isotropic case, in which the 2$s$ and 2$p$ states are degenerate. It will be interesting to probe the nondegenerate 2$p_x$ and 2$p_y$ states in few-layer BP, using two-photon PL spectroscopy.

**6.3. Near-Field Optical Study**

In previous studies, various techniques based on far-field optical spectroscopy are used to probe the electronic and optical properties of 2D BP.[42,43,57] In such studies, restricted by the diffraction limit, the spatial resolution is poor, typically in the order of micrometers. Thus, the obtained information is averaged over the size of several to tens of micrometers. As is well-known, wrinkles,[187] foldings,[188] edge-related defects[189] and local strain[190] are easily to be introduced in 2D materials. These nanoscale variations are expected to significantly modify the electronic structure and optical properties, and hence the device performance. Near-field optical spectroscopy with nanoscale resolution, is a very powerful tool in this aspect.[191] Therefore, it will be of great interest to probe the local properties of anisotropic 2D BP, using near-field optical spectroscopy, to fully uncover the intrinsic properties in the nanometer scale.

## References

[1]   K. S. Novoselov, A. K. Geim, S. V. Morozov, D. Jiang, Y. Zhang, S. V. Dubonos, I. V. Grigorieva, and A. A. Firsov, *Science* **2004**, *306*, 666.
[2]   D. N. Basov, M. M. Fogler, A. Lanzara, F. Wang, and Y. Zhang, *Rev. Mod. Phys.* **2014**, *86*, 959.
[3]   A. H. C. Neto, F. Guinea, N. M. R. Peres, K. S. Novoselov, and A. K. Geim, *Rev. Mod. Phys.* **2009**, *81*, 109.
[4]   A. K. Geim and K. S. Novoselov, *Nat. Mater.* **2007**, *6*, 183.
[5]   P. Avouris, *Nano Lett.* **2010**, *10*, 4285.
[6]   F. Bonaccorso, Z. Sun, T. Hasan, and A. C. Ferrari, *Nat. Photonics* **2010**, *4*, 611.
[7]   R. V. Gorbachev, I. Riaz, R. R. Nair, R. Jalil, L. Britnell, B. D. Belle, E. W. Hill, K. S. Novoselov, K. Watanabe, T. Taniguchi, A. K. Geim, and P. Blake, *Small* **2011**, *7*, 465.
[8]   C. Gong, L. Li, Z. Li, H. Ji, A. Stern, Y. Xia, T. Cao, W. Bao, C. Wang, Y. Wang, Z. Q. Qiu, R. J. Cava, S. G. Louie, J. Xia, and X. Zhang, *Nature* **2017**, *546*, 265.
[9]   B. Huang, G. Clark, E. Navarro-Moratalla, D. R. Klein, R. Cheng, K. L. Seyler, D. Zhong, E. Schmidgall, M. A. McGuire, D. H. Cobden, W. Yao, D. Xiao, P. Jarillo-Herrero, and X. Xu, *Nature* **2017**, *546*, 270.
[10]  X. Xi, L. Zhao, Z. Wang, H. Berger, L. Forro, J. Shan, and K. F. Mak, *Nat. Nanotechnol.* **2015**, *10*, 765.

[11]	D. L. Duong, G. Ryu, A. Hoyer, C. Lin, M. Burghard, and K. Kern, *ACS Nano* **2017**, *11*, 1034.

[12]	Y. Yu, L. Ma, P. Cai, R. Zhong, C. Ye, J. Shen, G. D. Gu, X. H. Chen, and Y. Zhang, *Nature* **2019**, *575*, 156.

[13]	S. Manzeli, D. Ovchinnikov, D. Pasquier, O. V. Yazyev, and A. Kis, *Nat. Rev. Mater.* **2017**, *2*, 17033.

[14]	G. R. Bhimanapati, Z. Lin, V. Meunier, Y. Jung, J. Cha, S. Das, D. Xiao, Y. Son, M. S. Strano, V. R. Cooper, L. Liang, S. G. Louie, E. Ringe, W. Zhou, S. S. Kim, R. R. Naik, B. G. Sumpter, H. Terrones, F. Xia, Y. Wang, J. Zhu, D. Akinwande, N. Alem, J. A. Schuller, R. E. Schaak, M. Terrones, and J. A. Robinson, *ACS Nano* **2015**, *9*, 11509.

[15]	A. Castellanos-Gomez, *J. Phys. Chem. Lett.* **2015**, *6*, 4280.

[16]	A. Castellanos-Gomez, *Nat. Photonics* **2016**, *10*, 202.

[17]	K. F. Mak and J. Shan, *Nat. Photonics* **2016**, *10*, 216.

[18]	F. Xia, H. Wang, D. Xiao, M. Dubey, and A. Ramasubramaniam, *Nat. Photonics* **2014**, *8*, 899.

[19]	A. Carvalho, M. Wang, X. Zhu, A. S. Rodin, H. Su, and A. H. C. Neto, *Nat. Rev. Mater.* **2016**, *1*, 1.

[20]	F. Xia, H. Wang, J. C. M. Hwang, A. H. C. Neto, and L. Yang, *Nat. Rev. Phys.* **2019**, *1*, 306.

[21]	S. Demirci, N. Avazlı, E. Durgun, and S. Cahangirov, *Phys. Rev. B* **2017**, *95*, 115409.

[22]	L. C. Gomes and A. Carvalho, *Phys. Rev. B* **2015**, *92*, 085406.

[23]	X. Ling, H. Wang, S. Huang, F. Xia, and M. S. Dresselhaus, *Proc. Natl. Acad. Sci. USA* **2015**, *112*, 4523.

[24]	L. Li, Y. Yu, G. J. Ye, Q. Ge, X. Ou, H. Wu, D. Feng, X. H. Chen, and Y. Zhang, *Nat. Nanotechnol.* **2014**, *9*, 372.

[25]	K. F. Mak, C. Lee, J. Hone, J. Shan, and T. F. Heinz, *Phys. Rev. Lett.* **2010**, *105*, 136805.

[26]	A. Splendiani, L. Sun, Y. Zhang, T. Li, J. Kim, C. Y. Chim, G. Galli, and F. Wang, *Nano Lett.* **2010**, *10*, 1271.

[27]	B. Radisavljevic, A. Radenovic, J. Brivio, V. Giacometti, and A. Kis, *Nat. Nanotechnol.* **2011**, *6*, 147.

[28]	W. Zhao, Z. Ghorannevis, L. Chu, M. Toh, C. Kloc, P. H. Tan, and G. Eda, *ACS Nano* **2013**, *7*, 791.

[29]	K. F. Mak, J. Shan, and T. F. Heinz, *Phys. Rev. Lett.* **2010**, *104*, 176404.

[30]	C. H. Lui, Z. Li, K. F. Mak, E. Cappelluti, and T. F. Heinz, *Nat. Phys.* **2011**, *7*, 944.

[31]	Y. Cao, V. Fatemi, S. Fang, K. Watanabe, T. Taniguchi, E. Kaxiras, and P. Jarillo-Herrero, *Nature* **2018**, *556*, 43.

[32]	Y. Cao, V. Fatemi, A. Demir, S. Fang, S. L. Tomarken, J. Y. Luo, J. D. Sanchez-Yamagishi, K. Watanabe, T. Taniguchi, E. Kaxiras, R. C. Ashoori, and P. Jarillo-Herrero, *Nature* **2018**, *556*, 80.

[33]	E. M. Alexeev, D. A. Ruiz-Tijerina, M. Danovich, M. J. Hamer, D. J. Terry, P. K. Nayak, S. Ahn, S. Pak, J. Lee, J. I. Sohn, M. R. Molas, M. Koperski, K. Watanabe, T. Taniguchi, K. S. Novoselov, R. V. Gorbachev, H. S. Shin, V. I. Fal'ko, and A. I. Tartakovskii, *Nature* **2019**, *567*, 81.

[34]	C. Jin, E. C. Regan, A. Yan, M. Iqbal Bakti Utama, D. Wang, S. Zhao, Y. Qin, S. Yang, Z. Zheng, S. Shi, K. Watanabe, T. Taniguchi, S. Tongay, A. Zettl, and F. Wang, *Nature* **2019**,

*567*, 76.

[35] K. L. Seyler, P. Rivera, H. Yu, N. P. Wilson, E. L. Ray, D. G. Mandrus, J. Yan, W. Yao, and X. Xu, *Nature* **2019**, *567*, 66.

[36] K. Tran, G. Moody, F. Wu, X. Lu, J. Choi, K. Kim, A. Rai, D. A. Sanchez, J. Quan, A. Singh, J. Embley, A. Zepeda, M. Campbell, T. Autry, T. Taniguchi, K. Watanabe, N. Lu, S. K. Banerjee, K. L. Silverman, S. Kim, E. Tutuc, L. Yang, A. H. MacDonald, and X. Li, *Nature* **2019**, *567*, 71.

[37] P. W. Bridgman, *J. Am. Chem. Soc.* **1914**, *36*, 1344.

[38] R. Hultgren, N. S. Gingrich, and B. E. Warren, *J. Chem. Phys.* **1935**, *3*, 351.

[39] H. Liu, A. T. Neal, Z. Zhu, Z. Luo, X. Xu, D. Tomának, and P. D. Ye, *ACS Nano* **2014**, *8*, 4033.

[40] F. Xia, H. Wang, and Y. Jia, *Nat. Commun.* **2014**, *5*, 4458.

[41] L. Peng, S. A. Wells, C. R. Ryder, M. C. Hersam, and M. Grayson, *Phys. Rev. Lett.* **2018**, *120*, 086801.

[42] G. Zhang, S. Huang, A. Chaves, C. Song, V. O. Ozcelik, T. Low, and H. Yan, *Nat. Commun.* **2017**, *8*, 14071.

[43] L. Li, J. Kim, C. Jin, G. J. Ye, D. Y. Qiu, F. H. da Jornada, Z. Shi, L. Chen, Z. Zhang, F. Yang, K. Watanabe, T. Taniguchi, W. Ren, S. G. Louie, X. H. Chen, Y. Zhang, and F. Wang, *Nat. Nanotechnol.* **2017**, *12*, 21.

[44] X. Wang, A. M. Jones, K. L. Seyler, V. Tran, Y. Jia, H. Zhao, H. Wang, L. Yang, X. Xu, and F. Xia, *Nat. Nanotechnol.* **2015**, *10*, 517.

[45] T. Low, A. S. Rodin, A. Carvalho, Y. Jiang, H. Wang, F. Xia, and A. H. C. Neto, *Phys. Rev. B* **2014**, *90*, 075434.

[46] N. Mao, J. Tang, L. Xie, J. Wu, B. Han, J. Lin, S. Deng, W. Ji, H. Xu, K. Liu, L. Tong, and J. Zhang, *J. Am. Chem. Soc.* **2016**, *138*, 300.

[47] S. Lee, F. Yang, J. Suh, S. Yang, Y. Lee, G. Li, H. Sung Choe, A. Suslu, Y. Chen, C. Ko, J. Park, K. Liu, J. Li, K. Hippalgaonkar, J. J. Urban, S. Tongay, and J. Wu, *Nat. Commun.* **2015**, *6*, 8573.

[48] Z. Luo, J. Maassen, Y. Deng, Y. Du, R. P. Garrelts, M. S. Lundstrom, P. D. Ye, and X. Xu, *Nat. Commun.* **2015**, *6*, 8572.

[49] H. Jang, J. D. Wood, C. R. Ryder, M. C. Hersam, and D. G. Cahill, *Adv. Mater.* **2015**, *27*, 8017.

[50] J. W. Jiang and H. S. Park, *Nat. Commun.* **2014**, *5*, 4727.

[51] Q. Wei and X. Peng, *Appl. Phys. Lett.* **2014**, *104*, 251915.

[52] J. Qiao, X. Kong, Z. X. Hu, F. Yang, and W. Ji, *Nat. Commun.* **2014**, *5*, 4475.

[53] A. N. Rudenko and M. I. Katsnelson, *Phys. Rev. B* **2014**, *89*, 201408(R).

[54] C. Chen, F. Chen, X. Chen, B. Deng, B. Eng, D. Jung, Q. Guo, S. Yuan, K. Watanabe, T. Taniguchi, M. L. Lee, and F. Xia, *Nano Lett.* **2019**, *19*, 1488.

[55] S. Zhang, J. Yang, R. Xu, F. Wang, W. Li, M. Ghufran, Y. W. Zhang, Z. Yu, G. Zhang, Q. Qin, and Y. Lu, *ACS Nano* **2014**, *8*, 9590.

[56] J. Yang, R. Xu, J. Pei, Y. W. Myint, F. Wang, Z. Wang, S. Zhang, Z. Yu, and Y. Lu, *Light-Sci. Appl.* **2015**, *4*, e312.

[57] G. Zhang, A. Chaves, S. Huang, F. Wang, Q. Xing, T. Low, and H. Yan, *Sci. Adv.* **2018**, *4*, eaap9977.

[58]    A. Chaves, T. Low, P. Avouris, D. Çakır, and F. M. Peeters, *Phys. Rev. B* **2015**, *91*, 155311.

[59]    A. Chaves, M. Z. Mayers, F. M. Peeters, and D. R. Reichman, *Phys. Rev. B* **2016**, *93*, 115314.

[60]    X. Chen, X. Lu, B. Deng, O. Sinai, Y. Shao, C. Li, S. Yuan, V. Tran, K. Watanabe, T. Taniguchi, D. Naveh, L. Yang, and F. Xia, *Nat. Commun.* **2017**, *8*, 1672.

[61]    C. Chen, X. Lu, B. Deng, X. Chen, Q. Guo, C. Li, C. Ma, S. Yuan, E. Sung, K. Watanabe, T. Taniguchi, L. Yang, and F. Xia, *Sci. Adv.* **2020**, *6*, eaay6134.

[62]    S. Huang, G. Zhang, F. Fan, C. Song, F. Wang, Q. Xing, C. Wang, H. Wu, and H. Yan, *Nat. Commun.* **2019**, *10*, 2447.

[63]    Z. Zhang, L. Li, J. Horng, N. Z. Wang, F. Yang, Y. Yu, Y. Zhang, G. Chen, K. Watanabe, T. Taniguchi, X. H. Chen, F. Wang, and Y. Zhang, *Nano Lett.* **2017**, *17*, 6097.

[64]    Q. Guo, A. Pospischil, M. Bhuiyan, H. Jiang, H. Tian, D. Farmer, B. Deng, C. Li, S. J. Han, H. Wang, Q. Xia, T. P. Ma, T. Mueller, and F. Xia, *Nano Lett.* **2016**, *16*, 4648.

[65]    A. Gao, J. Lai, Y. Wang, Z. Zhu, J. Zeng, G. Yu, N. Wang, W. Chen, T. Cao, W. Hu, D. Sun, X. Chen, F. Miao, Y. Shi, and X. Wang, *Nat. Nanotechnol.* **2019**, *14*, 217.

[66]    J. Bullock, M. Amani, J. Cho, Y. Z. Chen, G. H. Ahn, V. Adinolfi, V. R. Shrestha, Y. Gao, K. B. Crozier, Y. L. Chueh, and A. Javey, *Nat. Photonics* **2018**, *12*, 601.

[67]    R. J. Suess, E. Leong, J. L. Garrett, T. Zhou, R. Salem, J. N. Munday, T. E. Murphy, and M. Mittendorff, *2D Mater.* **2016**, *3*, 041006.

[68]    Y. Zhang, S. Wang, S. Chen, Q. Zhang, X. Wang, X. Zhu, X. Zhang, X. Xu, T. Yang, M. He, X. Yang, Z. Li, X. Chen, M. Wu, Y. Lu, R. Ma, W. Lu, and A. Pan, *Adv. Mater.* **2020**, *32*, e1808319.

[69]    R. W. Keyes, *Phys. Rev.* **1953**, *92*, 580.

[70]    D. Warschauer, *J. Appl. Phys.* **1963**, *34*, 1853.

[71]    Y. Takao and A. Morita, *Physica B* **1981**, *105B*, 93.

[72]    V. Tran, R. Soklaski, Y. Liang, and L. Yang, *Phys. Rev. B* **2014**, *89*, 235319.

[73]    A. S. Rodin, A. Carvalho, and A. H. C. Neto, *Phys. Rev. Lett.* **2014**, *112*, 176801.

[74]    S. P. Koenig, R. A. Doganov, H. Schmidt, A. H. C. Neto, and B. Özyilmaz, *Appl. Phys. Lett.* **2014**, *104*, 103106.

[75]    N. Ehlen, B. V. Senkovskiy, A. V. Fedorov, A. Perucchi, P. Di Pietro, A. Sanna, G. Profeta, L. Petaccia, and A. Grüneis, *Phys. Rev. B* **2016**, *94*, 245410.

[76]    B. Liao, H. Zhao, E. Najafi, X. Yan, H. Tian, J. Tice, A. J. Minnich, H. Wang, and A. H. Zewail, *Nano Lett.* **2017**, *17*, 3675.

[77]    J. He, D. He, Y. Wang, Q. Cui, M. Z. Bellus, H. Y. Chiu, and H. Zhao, *ACS Nano* **2015**, *9*, 6436.

[78]    G. Long, D. Maryenko, J. Shen, S. Xu, J. Hou, Z. Wu, W. K. Wong, T. Han, J. Lin, Y. Cai, R. Lortz, and N. Wang, *Nano Lett.* **2016**, *16*, 7768.

[79]    X. Chen, Y. Wu, Z. Wu, Y. Han, S. Xu, L. Wang, W. Ye, T. Han, Y. He, Y. Cai, and N. Wang, *Nat. Commun.* **2015**, *6*, 7315.

[80]    L. Li, G. J. Ye, V. Tran, R. Fei, G. Chen, H. Wang, J. Wang, K. Watanabe, T. Taniguchi, L. Yang, X. H. Chen, and Y. Zhang, *Nat. Nanotechnol.* **2015**, *10*, 608.

[81]    F. Yang, Z. Zhang, N. Z. Wang, G. J. Ye, W. Lou, X. Zhou, K. Watanabe, T. Taniguchi, K. Chang, X. H. Chen, and Y. Zhang, *Nano Lett.* **2018**, *18*, 6611.

[82]    L. Li, F. Yang, G. J. Ye, Z. Zhang, Z. Zhu, W. Lou, X. Zhou, L. Li, K. Watanabe, T. Taniguchi, K.

Chang, Y. Wang, X. H. Chen, and Y. Zhang, *Nat. Nanotechnol.* **2016**, *11*, 593.

[83] J. Yang, S. Tran, J. Wu, S. Che, P. Stepanov, T. Taniguchi, K. Watanabe, H. Baek, D. Smirnov, R. Chen, and C. N. Lau, *Nano Lett.* **2018**, *18*, 229.

[84] S. Tran, J. Yang, M. Gillgren, T. Espiritu, Y. Shi, K. Watanabe, T. Taniguchi, S. Moon, H. Baek, D. Smirnov, M. Bockrath, R. Chen, and C. N. Lau, *Sci. Adv.* **2017**, *3*, e1603179.

[85] A. N. Rudenko, S. Yuan, and M. I. Katsnelson, *Phys. Rev. B* **2015**, *92*, 085419.

[86] S. Dong, A. Zhang, K. Liu, J. Ji, Y. G. Ye, X. G. Luo, X. H. Chen, X. Ma, Y. Jie, C. Chen, X. Wang, and Q. Zhang, *Phys. Rev. Lett.* **2016**, *116*, 087401.

[87] Y. Liu, Z. Qiu, A. Carvalho, Y. Bao, H. Xu, S. J. Tan, W. Liu, A. H. C. Neto, K. P. Loh, and J. Lu, *Nano Lett.* **2017**, *17*, 1970.

[88] D. Li, H. Jussila, L. Karvonen, G. Ye, H. Lipsanen, X. Chen, and Z. Sun, *Sci. Rep.* **2015**, *5*, 15899.

[89] N. Youngblood, R. Peng, A. Nemilentsau, T. Low, and M. Li, *ACS Photonics* **2016**, *4*, 8.

[90] M. J. Rodrigues, C. J. de Matos, Y. W. Ho, H. Peixoto, R. E. de Oliveira, H. Y. Wu, A. H. C. Neto, and J. Viana-Gomes, *Adv. Mater.* **2016**, *28*, 10693.

[91] A. Autere, C. R. Ryder, A. Saynatjoki, L. Karvonen, B. Amirsolaimani, R. A. Norwood, N. Peyghambarian, K. Kieu, H. Lipsanen, M. C. Hersam, and Z. Sun, *J. Phys. Chem. Lett.* **2017**, *8*, 1343.

[92] A. C. Ferrari, J. C. Meyer, V. Scardaci, C. Casiraghi, M. Lazzeri, F. Mauri, S. Piscanec, D. Jiang, K. S. Novoselov, S. Roth, and A. K. Geim, *Phys. Rev. Lett.* **2006**, *97*, 187401.

[93] C. Lee, H. Yan, L. E. Brus, T. F. Heinz, and J. Hone, *ACS Nano* **2010**, *4*, 2695.

[94] H. B. Ribeiro, M. A. Pimenta, C. J. S. de Matos, R. L. Moreira, A. S. Rodin, J. D. Zapata, E. A. T. de Souza, and A. H. C. Neto, *ACS Nano* **2015**, *9*, 4270.

[95] J. Wu, N. Mao, L. Xie, H. Xu, and J. Zhang, *Angew. Chem. Int. Ed.* **2015**, *54*, 2366.

[96] A. Favron, F. A. Goudreault, V. Gosselin, J. Groulx, M. Cote, R. Leonelli, J. F. Germain, A. L. Phaneuf-L'Heureux, S. Francoeur, and R. Martel, *Nano Lett.* **2018**, *18*, 1018.

[97] X. Ling, S. Huang, E. H. Hasdeo, L. Liang, W. M. Parkin, Y. Tatsumi, A. R. Nugraha, A. A. Puretzky, P. M. Das, B. G. Sumpter, D. B. Geohegan, J. Kong, R. Saito, M. Drndic, V. Meunier, and M. S. Dresselhaus, *Nano Lett.* **2016**, *16*, 2260.

[98] J. Kim, J. U. Lee, J. Lee, H. J. Park, Z. Lee, C. Lee, and H. Cheong, *Nanoscale* **2015**, *7*, 18708.

[99] M. L. Lin, Y. C. Leng, X. Cong, D. Meng, J. Wang, X. L. Li, B. Yu, X. L. Liu, X. F. Yu, and P. H. Tan, *Sci. Bull.* **2020**, *65*, 1894.

[100] N. Mao, J. Wu, B. Han, J. Lin, L. Tong, and J. Zhang, *Small* **2016**, *12*, 2627.

[101] X. Miao, G. Zhang, F. Wang, H. Yan, and M. Ji, *Nano Lett.* **2018**, *18*, 3053.

[102] N. Mao, S. Zhang, J. Wu, J. Zhang, and L. Tong, *Small Methods* **2018**, *2*, 1700409.

[103] X. Li, B. Deng, X. Wang, S. Chen, M. Vaisman, S. Karato, G. Pan, M. L. Lee, J. Cha, H. Wang, and F. Xia, *2D Mater.* **2015**, *2*, 031002.

[104] C. Li, Y. Wu, B. Deng, Y. Xie, Q. Guo, S. Yuan, X. Chen, M. Bhuiyan, Z. Wu, K. Watanabe, T. Taniguchi, H. Wang, J. J. Cha, M. Snure, Y. Fei, and F. Xia, *Adv. Mater.* **2018**, *30*, 1703748.

[105] J. B. Smith, D. Hagaman, and H. F. Ji, *Nanotechnology* **2016**, *27*, 215602.

[106] A. Castellanos-Gomez, L. Vicarelli, E. Prada, J. O. Island, K. L. Narasimha-Acharya, S. I. Blanter, D. J. Groenendijk, M. Buscema, G. A. Steele, J. V. Alvarez, H. W. Zandbergen, J. J. Palacios, and H. S. J. van der Zant, *2D Mater.* **2014**, *1*, 025001.

[107] R. R. Nair, P. Blake, A. N. Grigorenko, K. S. Novoselov, T. J. Booth, T. Stauber, N. M. R. Peres, and A. K. Geim, *Science* **2008**, *320*, 1308.

[108] X. L. Li, W. P. Han, J. B. Wu, X. F. Qiao, J. Zhang, and P. H. Tan, *Adv. Funct. Mater.* **2017**, *27*, 1604468.

[109] K. F. Mak, M. Y. Sfeir, Y. Wu, C. H. Lui, J. A. Misewich, and T. F. Heinz, *Phys. Rev. Lett.* **2008**, *101*, 196405.

[110] H. Yan, X. Li, B. Chandra, G. Tulevski, Y. Wu, M. Freitag, W. Zhu, P. Avouris, and F. Xia, *Nat. Nanotechnol.* **2012**, *7*, 330.

[111] C. Lin, R. Grassi, T. Low, and A. S. Helmy, *Nano Lett.* **2016**, *16*, 1683.

[112] F. Wang, G. Dukovic, L. E. Brus, and T. F. Heinz, *Science* **2005**, *308*, 838.

[113] C. D. Spataru, S. Ismail-Beigi, L. X. Benedict, and S. G. Louie, *Phys. Rev. Lett.* **2004**, *92*, 077402.

[114] A. Chernikov, T. C. Berkelbach, H. M. Hill, A. Rigosi, Y. Li, O. B. Aslan, D. R. Reichman, M. S. Hybertsen, and T. F. Heinz, *Phys. Rev. Lett.* **2014**, *113*, 076802.

[115] K. He, N. Kumar, L. Zhao, Z. Wang, K. F. Mak, H. Zhao, and J. Shan, *Phys. Rev. Lett.* **2014**, *113*, 026803.

[116] Z. Ye, T. Cao, K. O'Brien, H. Zhu, X. Yin, Y. Wang, S. G. Louie, and X. Zhang, *Nature* **2014**, *513*, 214.

[117] M. M. Ugeda, A. J. Bradley, S. F. Shi, F. H. da Jornada, Y. Zhang, D. Y. Qiu, W. Ruan, S. K. Mo, Z. Hussain, Z. X. Shen, F. Wang, S. G. Louie, and M. F. Crommie, *Nat. Mater.* **2014**, *13*, 1091.

[118] P. Li and I. Appelbaum, *Phys. Rev. B* **2014**, *90*, 115439.

[119] V. Tran, R. Fei, and L. Yang, *2D Mater.* **2015**, *2*, 044014.

[120] C. Wang, G. Zhang, S. Huang, Y. Xie, and H. Yan, *Adv. Opt. Mater.* **2020**, *8*, 1900996.

[121] V. Wang, Y. C. Liu, Y. Kawazoe, and W. T. Geng, *J. Phys. Chem. Lett.* **2015**, *6*, 4876.

[122] K. Yamanaka, T. Fukunaga, N. Tsukada, K. L. I. Kobayashi, and M. Ishii, *Appl. Phys. Lett.* **1986**, *48*, 840.

[123] R. C. Miller, A. C. Gossard, G. D. Sanders, Y. C. Chang, and J. N. Schulman, *Phys. Rev. B* **1985**, *32*, 8452.

[124] W. S. Whitney, M. C. Sherrott, D. Jariwala, W. H. Lin, H. A. Bechtel, G. R. Rossman, and H. A. Atwater, *Nano Lett.* **2017**, *17*, 78.

[125] A. Favron, E. Gaufrès, F. Fossard, A. L. Phaneuf-L'Heureux, N. Y. Tang, P. L. Lévesque, A. Loiseau, R. Leonelli, S. Francoeur, and R. Martel, *Nat. Mater.* **2015**, *14*, 826.

[126] J. S. Kim, Y. Liu, W. Zhu, S. Kim, D. Wu, L. Tao, A. Dodabalapur, K. Lai, and D. Akinwande, *Sci. Rep.* **2015**, *5*, 8989.

[127] J. O. Island, G. A. Steele, H. S. J. van der Zant, and A. Castellanos-Gomez, *2D Mater.* **2015**, *2*, 011002.

[128] J. D. Wood, S. A. Wells, D. Jariwala, K. S. Chen, E. Cho, V. K. Sangwan, X. Liu, L. J. Lauhon, T. J. Marks, and M. C. Hersam, *Nano Lett.* **2014**, *14*, 6964.

[129] V. Artel, Q. Guo, H. Cohen, R. Gasper, A. Ramasubramaniam, F. Xia, and D. Naveh, *npj 2D Mater. Appl.* **2017**, *1*, 6.

[130] F. Wang, G. Zhang, S. Huang, C. Song, C. Wang, Q. Xing, Y. Lei, and H. Yan, *Phys. Rev. B* **2019**, *99*, 075427.

[131] W. Luo, R. Yang, J. Liu, Y. Zhao, W. Zhu, and G. M. Xia, *Nanotechnology* **2017**, *28*,

285301.

[132] Y. Li, Z. Hu, S. Lin, S. K. Lai, W. Ji, and S. P. Lau, *Adv. Funct. Mater.* **2017**, *27*, 1600986.

[133] A. Avsar, I. J. Vera-Marun, J. Y. Tan, K. Watanabe, T. Taniguchi, A. H. C. Neto, and B. Ozyilmaz, *ACS Nano* **2015**, *9*, 4138.

[134] R. A. Doganov, E. C. O'Farrell, S. P. Koenig, Y. Yeo, A. Ziletti, A. Carvalho, D. K. Campbell, D. F. Coker, K. Watanabe, T. Taniguchi, A. H. C. Neto, and B. Ozyilmaz, *Nat. Commun.* **2015**, *6*, 6647.

[135] D. He, Y. Wang, Y. Huang, Y. Shi, X. Wang, and X. Duan, *Nano Lett.* **2019**, *19*, 331.

[136] J. Pei, X. Gai, J. Yang, X. Wang, Z. Yu, D. Y. Choi, B. Luther-Davies, and Y. Lu, *Nat. Commun.* **2016**, *7*, 10450.

[137] S. Huang, F. Wang, G. Zhang, C. Song, Y. Lei, Q. Xing, C. Wang, Y. Zhang, J. Zhang, Y. Xie, L. Mu, C. Cong, M. Huang, and H. Yan, *Phys. Rev. Lett.* **2020**, *125*, 156802.

[138] J. Wang, A. Rousseau, M. Yang, T. Low, S. Francoeur, and S. Kena-Cohen, *Nano Lett.* **2020**, *20*, 3651.

[139] J. Kim, S. S. Baik, S. H. Ryu, Y. Sohn, S. Park, B. G. Park, J. Denlinger, Y. Yi, H. J. Choi, and K. S. Kim, *Science* **2015**, *349*, 723.

[140] B. Deng, V. Tran, Y. Xie, H. Jiang, C. Li, Q. Guo, X. Wang, H. Tian, S. J. Koester, H. Wang, J. J. Cha, Q. Xia, L. Yang, and F. Xia, *Nat. Commun.* **2017**, *8*, 14474.

[141] D. Li, J. R. Xu, K. Ba, N. Xuan, M. Chen, Z. Sun, Y. Z. Zhang, and Z. Zhang, *2D Mater.* **2017**, *4*, 031009.

[142] S. L. Yan, Z. J. Xie, J. H. Chen, T. Taniguchi, and K. Watanabe, *Chin. Phys. Lett.* **2017**, *34*, 047304.

[143] L. Ju, L. Wang, T. Cao, T. Taniguchi, K. Watanabe, S. G. Louie, F. Rana, J. Park, J. Hone, F. Wang, and P. L. McEuen, *Science* **2017**, *358*, 907.

[144] Y. Zhang, T. T. Tang, C. Girit, Z. Hao, M. C. Martin, A. Zettl, M. F. Crommie, Y. R. Shen, and F. Wang, *Nature* **2009**, *459*, 820.

[145] D. Çakır, H. Sahin, and F. M. Peeters, *Phys. Rev. B* **2014**, *90*, 205421.

[146] X. Peng, Q. Wei, and A. Copple, *Phys. Rev. B* **2014**, *90*, 085402.

[147] P. San-Jose, V. Parente, F. Guinea, R. Roldán, and E. Prada, *Phys. Rev. X* **2016**, *6*, 031046.

[148] J. Quereda, P. San-Jose, V. Parente, L. Vaquero-Garzon, A. J. Molina-Mendoza, N. Agrait, G. Rubio-Bollinger, F. Guinea, R. Roldan, and A. Castellanos-Gomez, *Nano Lett.* **2016**, *16*, 2931.

[149] H. J. Conley, B. Wang, J. I. Ziegler, R. F. Haglund Jr., S. T. Pantelides, and K. I. Bolotin, *Nano Lett.* **2013**, *13*, 3626.

[150] K. He, C. Poole, K. F. Mak, and J. Shan, *Nano Lett.* **2013**, *13*, 2931.

[151] C. R. Zhu, G. Wang, B. L. Liu, X. Marie, X. F. Qiao, X. Zhang, X. X. Wu, H. Fan, P. H. Tan, T. Amand, and B. Urbaszek, *Phys. Rev. B* **2013**, *88*, 121301(R).

[152] H. Duan, M. Yang, and R. Wang, *Physica E* **2016**, *81*, 177.

[153] E. Taghizadeh Sisakht, F. Fazileh, M. H. Zare, M. Zarenia, and F. M. Peeters, *Phys. Rev. B* **2016**, *94*, 085417.

[154] J. W. Jiang and H. S. Park, *Phys. Rev. B* **2015**, *91*, 235118.

[155] C. Wang, Q. He, U. Halim, Y. Liu, E. Zhu, Z. Lin, H. Xiao, X. Duan, Z. Feng, R. Cheng, N. O. Weiss, G. Ye, Y. C. Huang, H. Wu, H. C. Cheng, I. Shakir, L. Liao, X. Chen, W. A. Goddard, III, Y. Huang, and X. Duan, *Nature* **2018**, *555*, 231.

[156] S. Tongay, H. Sahin, C. Ko, A. Luce, W. Fan, K. Liu, J. Zhou, Y. S. Huang, C. H. Ho, J. Yan, D. F. Ogletree, S. Aloni, J. Ji, S. Li, J. Li, F. M. Peeters, and J. Wu, *Nat. Commun.* **2014**, *5*, 3252.

[157] W. Chen, J. Gu, Q. Liu, R. Luo, L. Yao, B. Sun, W. Zhang, H. Su, B. Chen, P. Liu, and D. Zhang, *ACS Nano* **2018**, *12*, 308.

[158] C. Jin, J. Kim, K. Wu, B. Chen, E. S. Barnard, J. Suh, Z. Shi, S. G. Drapcho, J. Wu, P. J. Schuck, S. Tongay, and F. Wang, *Adv. Funct. Mater.* **2017**, *27*, 1601741.

[159] Y. Li, A. Chernikov, X. Zhang, A. Rigosi, H. M. Hill, A. M. van der Zande, D. A. Chenet, E. M. Shih, J. Hone, and T. F. Heinz, *Phys. Rev. B* **2014**, *90*, 205422.

[160] A. M. Fox, *Optical properties of solids, second edition*, Oxford University Press, New York, **2010**.

[161] G. Wang, A. Chernikov, M. M. Glazov, T. F. Heinz, X. Marie, T. Amand, and B. Urbaszek, *Rev. Mod. Phys.* **2018**, *90*, 021001.

[162] T. Olsen, S. Latini, F. Rasmussen, and K. S. Thygesen, *Phys. Rev. Lett.* **2016**, *116*, 056401.

[163] T. Tian, D. Scullion, D. Hughes, L. H. Li, C. J. Shih, J. Coleman, M. Chhowalla, and E. J. G. Santos, *Nano Lett.* **2020**, *20*, 841.

[164] D. Unuchek, A. Ciarrocchi, A. Avsar, Z. Sun, K. Watanabe, T. Taniguchi, and A. Kis, *Nat. Nanotechnol.* **2019**, *14*, 1104.

[165] M. Kulig, J. Zipfel, P. Nagler, S. Blanter, C. Schuller, T. Korn, N. Paradiso, M. M. Glazov, and A. Chernikov, *Phys. Rev. Lett.* **2018**, *120*, 207401.

[166] X. X. Zhang, T. Cao, Z. Lu, Y. C. Lin, F. Zhang, Y. Wang, Z. Li, J. C. Hone, J. A. Robinson, D. Smirnov, S. G. Louie, and T. F. Heinz, *Nat. Nanotechnol.* **2017**, *12*, 883.

[167] X. X. Zhang, Y. You, S. Y. Zhao, and T. F. Heinz, *Phys. Rev. Lett.* **2015**, *115*, 257403.

[168] J. S. Ross, S. Wu, H. Yu, N. J. Ghimire, A. M. Jones, G. Aivazian, J. Yan, D. G. Mandrus, D. Xiao, W. Yao, and X. Xu, *Nat. Commun.* **2013**, *4*, 1474.

[169] K. F. Mak, K. He, C. Lee, G. H. Lee, J. Hone, T. F. Heinz, and J. Shan, *Nat. Mater.* **2013**, *12*, 207.

[170] Y. You, X. X. Zhang, T. C. Berkelbach, M. S. Hybertsen, D. R. Reichman, and T. F. Heinz, *Nat. Phys.* **2015**, *11*, 477.

[171] A. Chernikov, A. M. van der Zande, H. M. Hill, A. F. Rigosi, A. Velauthapillai, J. Hone, and T. F. Heinz, *Phys. Rev. Lett.* **2015**, *115*, 126802.

[172] M. Goryca, J. Li, A. V. Stier, T. Taniguchi, K. Watanabe, E. Courtade, S. Shree, C. Robert, B. Urbaszek, X. Marie, and S. A. Crooker, *Nat. Commun.* **2019**, *10*, 4172.

[173] A. V. Stier, N. P. Wilson, K. A. Velizhanin, J. Kono, X. Xu, and S. A. Crooker, *Phys. Rev. Lett.* **2018**, *120*, 057405.

[174] X. Liu, T. Galfsky, Z. Sun, F. Xia, E. C. Lin, Y. H. Lee, S. Kéna-Cohen, and V. M. Menon, *Nat. Photonics* **2014**, *9*, 30.

[175] Y. G. Ye, Z. J. Wong, X. Lu, X. Ni, H. Zhu, X. Chen, Y. Wang, and X. Zhang, *Nat. Photonics* **2015**, *9*, 733.

[176] D. Y. Qiu, F. H. da Jornada, and S. G. Louie, *Nano Lett.* **2017**, *17*, 4706.

[177] A. S. Rodin, A. Carvalho, and A. H. C. Neto, *Phys. Rev. B* **2014**, *90*, 075429.

[178] S. Arra, R. Babar, and M. Kabir, *Phys. Rev. B* **2019**, *99*, 045432.

[179] J. C. G. Henriques and N. M. R. Peres, *Phys. Rev. B* **2020**, *101*, 035406.

[180] P. Cudazzo, I. V. Tokatly, and A. Rubio, *Phys. Rev. B* **2011**, *84*, 085406.

[181] T. C. Berkelbach, M. S. Hybertsen, and D. R. Reichman, *Phys. Rev. B* **2013**, *88*, 045318.

[182] G. Zhang, S. Huang, F. Wang, Q. Xing, C. Song, C. Wang, Y. Lei, M. Huang, and H. Yan, *Nat. Commun.* **2020**, *11*, 1847.

[183] W. T. Masselink, P. J. Pearah, J. Klem, C. K. Peng, H. Morkoc, G. D. Sanders, and Y. C. Chang, *Phys. Rev. B* **1985**, *32*, 8027.

[184] Y. Masumoto, M. Matsuura, S. Tarucha, and H. Okamoto, *Phys. Rev. B* **1985**, *32*, 4275.

[185] T. Stauber, D. Noriega-Pérez, and J. Schliemann, *Phys. Rev. B* **2015**, *91*, 115407.

[186] S. Zhao, E. Wang, E. A. Üzer, S. Guo, K. Watanabe, T. Taniguchi, T. Nilges, Y. Zhang, B. Liu, X. Zou, and F. Wang, *Preprint at arXiv:1912.03644* **2019**.

[187] W. Zhu, T. Low, V. Perebeinos, A. A. Bol, Y. Zhu, H. Yan, J. Tersoff, and P. Avouris, *Nano Lett.* **2012**, *12*, 3431.

[188] T. Jiang, H. Liu, D. Huang, S. Zhang, Y. Li, X. Gong, Y. R. Shen, W. T. Liu, and S. Wu, *Nat. Nanotechnol.* **2014**, *9*, 825.

[189] T. X. Huang, X. Cong, S. S. Wu, K. Q. Lin, X. Yao, Y. H. He, J. B. Wu, Y. F. Bao, S. C. Huang, X. Wang, P. H. Tan, and B. Ren, *Nat. Commun.* **2019**, *10*, 5544.

[190] B. Lyu, H. Li, L. Jiang, W. Shan, C. Hu, A. Deng, Z. Ying, L. Wang, Y. Zhang, H. A. Bechtel, M. C. Martin, T. Taniguchi, K. Watanabe, W. Luo, F. Wang, and Z. Shi, *Nano Lett.* **2019**, *19*, 1982.

[191] A. Woessner, P. Alonso-Gonzalez, M. B. Lundeberg, Y. Gao, J. E. Barrios-Vargas, G. Navickaite, Q. Ma, D. Janner, K. Watanabe, A. W. Cummings, T. Taniguchi, V. Pruneri, S. Roche, P. Jarillo-Herrero, J. Hone, R. Hillenbrand, and F. H. Koppens, *Nat. Commun.* **2016**, *7*, 10783.